\documentclass{article}
\usepackage{graphicx}
\usepackage{amsmath}
\usepackage{amsfonts}
\usepackage{amssymb}

\begin{document}

\title{\textbf{Kontsevich-Witten Model From 2+1 Gravity: New} \textbf{Exact
Combinatorial Solution}}
\author{A. Kholodenko\thanks{\textit{E-mail address}: string@clemson.edu (A.L.Kholodenko)}\\375 H.L.Hunter Laboratories, \\Clemson University, Clemson, \\SC 29634-0973, USA}
\maketitle
\begin{abstract}
In previous publications ( J. Geom.Phys.38 (2001) 81-139 \ and references
therein ) the partition function for 2+1 gravity was constructed for the fixed
genus Riemann surface.With help of this function the dynamical transition from
pseudo-Anosov to periodic (Seifert-fibered) regime was studied. In this paper
the periodic regime is studied in some detail in order to recover major
results of Kontsevich (Comm.Math.Phys. 147 (1992) 1-23 ) inspired by earlier
work of Witten on topological two dimensional quantum gravity.To achieve this
goal some results from enumerative combinatorics have been used. The logical
developments are extensively illustrated using geometrically convincing
figures. This feature is helpful for development of some non traditional
applications (mentioned through the entire text) of obtained results to fields
other than theoretical particle physics.

MCS:83C45

\textit{Subj.Class}.:Quantum gravity

\end{abstract}

\section{Introduction}

\subsection{\noindent Motivation}

It may not be exaggeration to say that dynamics of 2+1 gravity is just an
interpretation of Nielsen -Thurston theory of surface homeomorphisms in terms
of concepts known in physics [1-4]. Apparently, the\ reverse task \ of
enriching mathematics with concepts known from physics had been accomplished
by Witten [5]. His physical intuition had revolutionized \ combinatorial
methods of algebraic geometry related to intersection theory \ on moduli space
of curves [6-8]. In his famous paper [9] Kontsevich had provided needed
mathematical justification to Witten's work. Since in both papers the initial
object of study is 2 dimensional quantum gravity it is only natural to expect
that the results of Witten and Kontsevich can be reobtained from more general
2+1 gravity model. The purpose of this paper is to demonstrate that this is
indeed possible.\ \ 

It should be noted that the results of Witten and Kontsevich (W-K) rely
heavily on the fact (proven by Kontsevich [9]) that the partition function of
2d gravity happens to be $\ \tau$ function of the Korteweg-de Vries (KdV)
hierarchy which is just a special case of \ more general
Kadomtsev-Petviashvili (KP) hierarchy of nonlinear exactly integrable
\ partial differential equations. Attempts to connect string theory with
exactly integrable systems had been made earlier. Good summary of earlier
efforts can be found in Ref.[10]. Recently, KP equations had been discussed in
relation to dynamics of 2+1 gravity [11]. The content of Ref.[11] apparently
has no connections with results of W-K. This does not exclude the possibility
that such connection might exist and requires further study. In the present
work we establish such connection (reduction) by reobtaining W-K results from
formalism developed earlier for 2+1 gravity [1-4]. This formalism is based on
some properties of Riemann surfaces. From algebraic geometry it is known that
every Riemann surface $\mathcal{S}$ can be described in terms of the
corresponding \ complex algebraic curve $\mathcal{C}$ [12] so that one can use
interchangeably algebro-geometric and complex analytic language to discuss the
same object, the Riemann surface. This,unfortunately, is not an easy task as
was noticed by Looijenga [7]. The existing difference in terminology is
especially apparent when one is interested in the issue of compactification of
the moduli space of Riemann surfaces (e.g. see Eq.(1.1) and related discussion)

Moduli spaces $\mathcal{M}_{g}$ are connected with Teichm\"{u}ller spaces
$\mathcal{T}_{g}$ of Riemann surfaces of genus $g\geq2$ in such a way that
$\mathcal{M}_{g}=$ $\mathcal{T}_{g}/\mathcal{\Gamma}$ with $\Gamma$ being the
mapping class group of $\mathcal{S}$. Both $\mathcal{M}_{g}$ and
${}\mathcal{T}_{g}$ have been discussed extensively in our earlier published
papers on 2+1 gravity [1-4]. The concept of moduli had been initially proposed
by Riemann \ who stated that isomorphism classes of closed Riemann surfaces of
genus $g\geq2$ are parametrized by 3g-3 complex parameters (or by 6g-6 real
parameters). The space of allowed values of these parameters is effectively
the moduli space $\mathcal{M}_{g}$ . More accurately, such moduli space is
called \textit{coarse moduli space}. It is complex analytic space of dimension
3g-3 but \textbf{not} a complex manifold. Presence of singularities (to be
discussed below) prevents this manifold from becoming a complex manifold
[6,13]. One way of describing both $\mathcal{M}_{g}$ and $\mathcal{T}_{g}$ is
through introduction of marking of $\mathcal{S}$ (accordingly of $\mathcal{C}$
). Marking of $\mathcal{S}$ can be made by introducing some $n$ boundary
components such that each of them is conformally equivalent to a punctured
disk [14]. Accordingly, for algebraic curves we select $n$
\textit{distinguished} (smooth) points on the curve. Smooth means that the
points are \textbf{not} located at the possible singularities of
\ $\mathcal{C}$. According to Deligne and Mumford [15] such singularities are
regular double points which, in the case of \ traditional visualization of
\ Riemann surfaces as \ closed surfaces with $g$ holes, are associated with
formation of nodes as depicted in Fig.1.

\bigskip%
\begin{figure}
[ptb]
\begin{center}
\includegraphics[
natheight=4.958000in,
natwidth=8.000400in,
height=2.3938in,
width=3.8467in
]%
{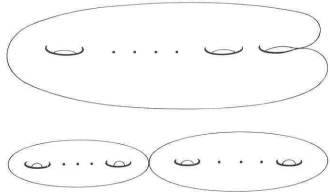}%
\caption{Various types of degeneration of the Riemann surfaces}%
\label{riemann}%
\end{center}
\end{figure}

The compactification $\mathcal{\bar{M}}_{g,n}$ \ of moduli space is achieved
by inclusion into $\mathcal{M}_{g,n}$ domains in parameter space associated
with Riemann surfaces with nodes. In the language of algebraic geometry
\ complex curves associated with such surfaces are called \textit{stable}.
\ Marked curves possess only finitely many authomorphisms. It is known [6]
that no stable curve can have more than 3g-3 nodes. This fact may be easily
understood if we recall that each Riemann surface admits pants decomposition
[2] so that 3g-3 nodes can be associated with degenerate geodesics whose
lengths are squeezed to zero. For Riemann surface $\mathcal{S}$ instead of (or
in addition to) marked points one can think about families of closed curves (
lamination sets [4]) and their intersections. By \ properly defining
intersection number(s) one can measure the nontriviality of such sets [6]. In
algebraic geometry such numbers are associated with cohomology classes known
also as $\mathit{tautological}$ \textit{classes} $\psi_{i}$ in $H^{2}%
(\mathcal{\bar{M}}_{g,n},\mathbf{Q})$. These numbers should \ not be confused
with \textit{geometrical} intersection numbers introduced by Thurston [16,17].
Roughly speaking such number $i$ measures the number of intersections between
a selected simple nontrivial closed curve $\alpha$ on $\mathcal{S}$ and
measured foliation $\mathcal{F}$: $i=i(\mathcal{F},\alpha)$.This observation
allows one to introduce equivalence relations between measured foliations. In
particular, foliations $\mathcal{F}_{1}$ and $\mathcal{F}_{2}$ are
\textit{projectively} equivalent if there is number $\lambda\in\mathbf{R}_{+}$
such that \ $i$($\mathcal{F}_{1}$,$\alpha)=\lambda i(\mathcal{F}_{2},\alpha).$
Following Thurston [17]and others [18] , we define the space of projective
laminations $\mathcal{PL}$($\mathcal{S}$) \ as space of equivalence classes of
measured foliations. Such defined space forms a \textit{boundary} of the
Teichm\"{u}ller space $\mathcal{T}_{g,n}.$ The compactification of the
Teichm\"{u}ller space $\mathcal{\bar{T}}$($\mathcal{S}$) is defined now as
\begin{equation}
\mathcal{\bar{T}}(\mathcal{S})=\mathcal{T}(\mathcal{S})\cup\mathcal{PL}%
(\mathcal{S}). \tag{1.1}%
\end{equation}
The set $\mathcal{\bar{T}}(\mathcal{S})$ is closed ball of real dimension
$6g-6+2n$ whose boundary is identified with $\mathcal{PL}$($\mathcal{S}$).
Since the moduli space is the quotient of $\mathcal{T(S)}$ by the mapping
class group $\Gamma$ , is clear that such defined compactification of the
Teichm\"{u}ller space leads accordingly to the compactification of the moduli
space. Such compactificatioon is apparently different from that of
Deligne-Mumford. The difference has some physical meaning. Indeed, in the
traditional, algebro-geometric case, no dynamics is involved while in the
Thurston's case the dynamics (that is ''time evolution'' ) is associated with
dynamically formed 3-manifolds, e.g. see chapters 8 and 9 of Thurston's
lecture notes [16]. These dynamical features are explained in physical terms
and illustrated in our recently published paper on statistical mechanics of
2+1 gravity [4] .

Since in 2 dimensional quantum gravity there is no time evolution, naturally,
there is no dynamical observables and theory is essentially topological.
Witten [5] had suggested to study the topological averages of the type
\begin{equation}
<\tau_{k_{1}}\cdot\cdot\cdot\tau_{k_{n}}>_{g}\equiv\int
\nolimits_{\mathcal{\bar{M}}_{g,n}}\psi_{1}^{k_{1}}\cdot\cdot\cdot\psi
_{n}^{k_{n}} \tag{1.2}%
\end{equation}
provided that $3g-3+n=\sum\nolimits_{i=1}^{n}k_{i}$ . To understand the
meaning of such an average it is \ instructive to consider the first
nontrivial example: $g=1,n=1$. In this case we obtain,
\begin{equation}
<\tau_{1}>_{1}=\int\nolimits_{\mathcal{\bar{M}}_{1,1}}\psi. \tag{1.3}%
\end{equation}
Witten had demonstrated that $<\tau_{1}>_{1}=\frac{1}{24}$ . This result had
been actually obtained earlier by Wolpert [19] who noticed that
\begin{equation}
<\tau_{1}>_{1}\dot{=}\frac{1}{2\pi^{2}}\int\nolimits_{\mathcal{\bar{M}}_{1,1}%
}\omega_{w-p} \tag{1.4}%
\end{equation}
where $\omega_{w-p}$ is familiar Weil-Petersson K\"{a}hler (1,1) form which
plays very important role in both the theory of Teichm\"{u}ller spaces [20]
and in string theory [21]. The dot above the equality sign has the following
meaning. The integral $\int\nolimits_{\mathcal{\bar{M}}_{1,1}}\omega_{w-p}$
had been independently calculated by Wolpert [22] and Penner [23] and is known
to be equal to $\dfrac{\pi^{2}}{6}$ so that the right hand side of Eq.(1.4) is
equal to $\frac{1}{12}.$ The same result was obtained by Witten [5] who argues
that it should be additionally divided by 2 because ''the generic elliptic
curve has two symmetries''. \ Thus, evidently, computation of ''tautological
averages'', Eq.(1.2), is effectively equivalent (up to numerical prefactors)
to calculation of \ the Weil-Petersson volumes of $\mathcal{\bar{M}}_{g,n}$.
Zograf [24] had calculated such volumes for $n\geq3$ punctured spheres. Based
on these results, Matone [25] had developed and solved non pertubatively 2d
Liouvillian gravity. Some of his results had been used and extended in the
work by Kaufman, Manin and Zagier [26]. These results had been further
streamlined in recently published monograph by Manin [27].

\subsection{From Chern classes to Grassmannians}

To make an additional connection with physics, we would like to reanalyze just
obtained results from the point of view of differential geometry of complex
manifolds [21,28]. In this regard, a closed \textit{integral} (1,1) form
$\omega$ \ on \ a complex manifold $M$ determines the equivalence class of a
line bundle $\lambda.$ Recall \ [21] that a line bundle $\lambda$ on manifold
\textit{M }is\textit{\ }an assignment of one\textit{\ }%
dimensional\textit{\ \ }complex \textit{vector} space $\lambda_{z}$ to each
point $z\in M$ . Sections of $\lambda_{z}$ are some functions $f_{\alpha}$
($z\in\mathcal{B}_{\alpha})$ assigning an element of $\lambda_{z}$ to each
point $z$. Vector spaces at different points should fit together and this
leads naturally to the fiber bundle constructions made of transition functions
$\phi_{\alpha_{\beta}}$ between different coordinate charts $\mathcal{B}%
_{\alpha}$. A metric on $\lambda_{z}$ \ is a set of positive functions
$g_{\alpha}$ such that $g_{\alpha}=\left|  \phi_{\alpha_{\beta}}\right|
^{2}g_{\beta}$. The covariant derivative is defined now as
\begin{equation}
\nabla_{j}f_{\alpha}=\{\frac{\partial}{\partial z_{j}}+\frac{\partial
}{\partial z_{j}}\ln g_{\alpha}\}f_{\alpha}. \tag{1.5}%
\end{equation}
The bundle $\lambda$ is \textit{holomorphic} if in addition
\begin{equation}
\bar{\nabla}_{j}f_{\alpha}=\frac{\partial}{\partial\bar{z}_{j}}f_{\alpha
}\text{ .} \tag{1.6}%
\end{equation}
With help of such defined covariant derivative the curvature tensor $F$ of
$\lambda$ is obtained in a standard way [28] with the result:
\begin{equation}
F=\frac{\partial^{2}}{\partial z\partial\bar{z}}\ln g_{\alpha}. \tag{1.7}%
\end{equation}
Finally, it can be shown [21,28], that
\begin{equation}
\omega=\frac{i}{2\pi}Fdz\wedge d\bar{z} \tag{1.8}%
\end{equation}
and that, actually, such defined (1,1) form coincides with the first Chern
class form $c_{1}(\lambda)$. In our case $M=$ $\mathcal{\bar{M}}_{g,n}$ and
the line bundle is actually a cotangent bundle $T^{\ast}(\mathcal{\bar{M}%
}_{g,n})$ made of quadratic differentials [Wolpert, positive] .Wolpert [29,30]
\ had demonstrated that
\begin{equation}
\frac{1}{\pi^{2}}\omega_{w-p}=\frac{i}{2\pi}F=c_{1}(\lambda). \tag{1.9}%
\end{equation}
Substituting this result into Eq.(1.4) we obtain,
\begin{equation}
<\tau_{1}>_{1}=\int\nolimits_{\mathcal{\bar{M}}_{1,1}}c_{1p}(\lambda),
\tag{1.10}%
\end{equation}
where $c_{1p}(\lambda)$=$c_{1}(\lambda)/2$ . In \ such notations this result
coincides with Eq.(1.7) of Witten's paper [5] (for $d=1$). The reason of
introducing the factor of 1/2 can be explained as follows. According to Wells,
Ref.[28], Ch-r 6, the traditional construction of Hodge line bundles for
manifolds requires that
\begin{equation}
\int\omega=\operatorname{integer}=c. \tag{1.11}%
\end{equation}
Moreover, if $c$ is an integer, it is actually equal to the Euler
characteristic of the manifold. In our case, $\mathcal{\bar{M}}_{1,1}$ is an
orbifold and for the orbifolds the Euler number can be rational number
according to Thurston [16]. Wells explains how to construct the Hodge bundle
for any K\"{a}hlerian manifold/orbifold. To this purpose, if the constant $c$
in Eq.(1.11) is not an integer, it is sufficient to rescale the form
$\omega:\Omega=c^{-1}\omega$ , so that the rescaled form \textbf{is} the Hodge
form. Since the moduli space $\mathcal{\bar{M}}_{g,n}$ is an orbifold,
Eq.(1.10) is acceptable and, hence, with such remarks, coincides with Eq.(1.7)
of Witten's paper (for $d=1$), Ref.[5].

Obtained results possess additional \ very important physical information
allowing us to make direct \ and simple connections with Grassmannians and,
hence, with exactly integrable systems, using arguments different from that of
Witten [5] and Kontsevich [9].

Let $G^{\mathbf{C}}(d,n)$ be complex Grassmann manifold. It \ is known that
\begin{equation}
G^{\mathbf{C}}(d,n)=\frac{U(d+n)}{U(d)U(n)}. \tag{1.12}%
\end{equation}
where $U(n)$ belongs to the unitary \ Lie group of $n\times n$ matrices.
Following Ref.[31,32] consider now the \textit{classifying} space $BU(n)$
defined by
\begin{equation}
BU(d)=\lim_{n\rightarrow\infty}\frac{U(d+n)}{U(d)U(n)}. \tag{1.13}%
\end{equation}
The name \textit{classifying} comes from the following observation. Suppose
there is a map $f$ from a base space $M$ to $BU(d)$, i.e. $f$: $M\rightarrow
BU(d)$. Construction of such a map for the case of orbifolds has been
developed by Baily [33] and, in particular, for $\mathcal{\bar{M}}_{g,n}$ by
Wolpert [29]. The $n$-vector complex bundle $\lambda$ over the base space $M$
can be expressed as a pullback of the \textit{standard} vector bundle $\xi$
over $BU(d)$ . That is $\lambda=f^{\ast}(\xi).$ The i-th Chern class
$c_{i}^{\ast\text{ }}$ of the vector bundle $\lambda$ can then be calculated
simply by the pullback $f^{\ast}(c_{i})$ $\in H^{2i}(M,R)$ of the element
$c_{i}$ in the 2i-th cohomology group of $BU(d).$ $G^{\mathbf{C}}(d,n)$ can be
embedded in the complex projective space. This can be achieved for any $n$ and
the limiting case $n\rightarrow\infty$ is known as\textit{\ Sato Grassmannian}
in the theory of KP equations [34]. We would like now to describe such an
embedding in some detail.

Recall that the complex projective space \textbf{P}$^{n}$ is defined as
\textbf{P}$^{n}:=(\mathbf{C}^{n+1}-\{\mathbf{0}\})/\sim$ with equivalence
relation $\sim$ defined as follows. If a point $P\in$\textbf{P}$^{n}$ is given
by an $n+1$ tuple ($z(0),...,z(n))$, then another $(n+1)$ tuple $(z\prime
(0),...,z\prime(n))$ defines \textbf{the same} point \ $P\in$\textbf{P}$^{n}$
if there is a \ nonzero number $c$ such that $z(i)=cz\prime(i),$ $\forall i$ ,
$i=0-n.$The system of standard coordinates ($U_{i}$ ,$\varphi_{i})$ enables us
to define a manifold structure on \textbf{P}$^{n}$ :
\[
U_{i}:=\{(z(0),...,z(n))\mid z(i)\neq0\},i=1-n
\]
and
\begin{equation}
\varphi_{i}:U_{i}\rightarrow\mathbf{C}^{n}:=\text{ (}\frac{z(0)}%
{z(i)},...,\frac{z(i-1)}{z(i)},\frac{z(i+1)}{z(i)},...,\frac{z(n)}{z(i)}).
\tag{1.14}%
\end{equation}
With help of such map $U_{i\text{ }}$ can be identified with the affine space
\textbf{C}$^{n}$ and the space \textbf{P}$^{n}$ can be made from different
coordinate patches $U_{i}$ with help of transition \ functions $\phi
_{\alpha_{\beta}}$ as discussed before.\ A \textbf{linear} space $\mathcal{L}$
in \textbf{P}$^{n}$ is defined as the set of points $P$=$(z(0),...,z(n))$ of
\textbf{P}$^{n}$ whose coordinates satisfy a system of linear equations
\begin{equation}
\sum\nolimits_{j=0}^{n}b_{aj}z(j)=0 \tag{1.15}%
\end{equation}
$\alpha=1,...,(n-d).$ The space $\mathcal{L}$ is $d$-dimensional if
$(n-d)\times(n+1)$ matrix of coefficients [$b_{aj}]$ has a nonzero
$(n-d)\times(n-d)$ minor. In this case there are $d+1$ points $P_{i}=$
($z_{i}(0),...,z_{i}(n))$ in $\mathcal{L}$ ($i=0-d)$ which span $\mathcal{L}$
. Naturally, $\mathcal{L}$ is a line if $d=1$, a plane if $d=2$ and a
hyperplane if $d>2$. We shall call these planes as $d$-planes following [35] .
$d$-planes in \textbf{P}$^{n}$ can be represented by the points in the
projective space \textbf{P}$^{N}$ whose dimension $N$ is given by
\begin{equation}
N=\frac{\left(  n+1\right)  !}{\left(  d+1\right)  !(n-d)!}-1. \tag{1.16}%
\end{equation}
To this purpose, let us fix a $d$-plane in \textbf{P}$^{n}$ and pick $d+1$
points $P_{i}=$($z_{i}(0),...,z_{i}(n))$ which span $\mathcal{L}$. Using these
points let us form $(d+1)\times(n+1)$ matrix [$p_{i}(j)]$ with $0\leq i\leq d$
and 0$\leq j\leq n.$ Let $j_{0},...,j_{d}$ be a sequence of integers with
0$\leq$ $j_{\beta}$ $\leq n$ and let $p(j_{0},...,j_{d})$ denote the
determinant of $(d+1)\times(d+1)$ matrix $[p_{i}(j_{\beta})]$ with $i$%
,$\beta=0,...,d.$ There will be $N$+1 determinants of such type and at least
one of them is nonzero by requirements of linear algebra. Hence, in view of
Eq.(1.14), we conclude that we can use these determinants to determine a point
in the complex projective space \textbf{P}$^{N}.$ The coordinates of this
point are called \textbf{Pl\"{u}cker} \textbf{coordinates} of $\mathcal{L}$ in
\textbf{P}$^{N}$ and such \ an embedding of the complex Grassmannian manifold
(of $d$-planes in \textbf{P}$^{n}$ space) into complex projective space
\textbf{P}$^{N}$ is called \textbf{Pl\"{u}cker embedding}. Not every point in
\textbf{P}$^{N}$ arises from $d$-plane in \textbf{P}$^{n}.$ Pl\"{u}cker
coordinates $p$($j_{0},...,j_{d})$ obey the following set of \ (Pl\"{u}cker)
equations
\begin{equation}
\sum\nolimits_{j=0}^{d+1}(-1)^{j}p(j_{0},...,j_{d-1}k_{j})p(k_{0}%
,...,\check{k}_{j},...,k_{d+1})=0. \tag{1.17}%
\end{equation}
Here $j_{0},...,j_{d-1}$ and $k_{0},.,k_{d+1}$are sequences of \ integers with
0$\leq j_{\beta}$ , $k_{\xi}\leq n$ with $\check{k}_{j}$ meaning that the
integer $k_{j}$ has been removed from the sequence.

As it is shown by Miwa et al [34] Pl\"{u}cker coordinates represent \ the
location of tau function of the KP hierarchy inside the Grassmannian while
Pl\"{u}cker equations are in one to one correspondence with the Hirota
bilinear equations. Hence, the connection between the averages given by
Eq.(1.2) and KP (or, more exactly, KdV) hierarchy naturally follows. The
determinants, which are points in \textbf{P}$^{N},$ have probabilistic meaning
which is discussed below. To this purpose we have to introduce some additional concepts.

\subsection{ From Grassmannians to Schubert varieties}

Consider a sequence of subspaces (cellular decomposition), i.e. $A_{0}\subset
A_{1}\subset...\subset A_{d}$ of a fixed space, e.g. \textbf{P}$^{n},$ each
properly contained in the next, whose dimensions dim $A_{i}=a_{i}$ , provided
that $0\leq a_{0}<a_{1}<...<a_{d}\leq n$ . Such construction is called
\textbf{flag}. Let now $\Omega(A_{0},...,A_{d})$ be the subset $\mathcal{L}$
of $G^{\mathbf{C}}(d,n)$ consisting of all $d$-planes satisfying $\dim$
$(\mathcal{L\cap}A_{i})\geq i$ for $i$=$0,...,d.$ Thus, if $\dim$ $A_{i}=i$
$\forall i$ , then $\Omega(A_{0},...,A_{d})$ is made of a single $d$-plane,
while if $\dim A_{i}=n-d+i$ $\forall i$ , then $\Omega(A_{0},...,A_{d}%
)=G^{\mathbf{C}}(d,n).$

\textbf{Definition 1.1.} $\Omega(A_{0},...,A_{d})$ is called \textit{Schubert
variety} corresponding to the flag $A_{0}\subset A_{1}\subset...\subset A_{d}$
. $\Omega(A_{0},...,A_{d})$ defines a homology (actually, cellular homology
[32] ) class in the homology ring $H_{\ast}^{{}}$($G^{\mathbf{C}%
}(d,n);\mathbf{Z}).$

\textbf{Definition 1.2.}Homology class in $H_{\ast}^{{}}$($G^{\mathbf{C}%
}(d,n);\mathbf{Z})$ is called \textit{Schubert cycle}.Because the homology
class depends only on the integers $a_{i}=$dim$A_{i}$ , it is appropriate to
write $\Omega(A_{0},...,A_{d})=\Omega(a_{0},...a_{d})$ where 0$\leq
a_{0}<a_{1}<...<a_{d}\leq n.$

It can be shown [12] that the product of any two Schubert cycles can be
uniquely expressed as a linear combination of other Schubert cycles.This
observation is central for development of \textit{Schubert calculus} [12,35,36].

\textbf{Remark 1.3}. Gepner [37] \ had demonstrated that \ all \ results of
rational conformal field theories can be \ actually obtained from physically
reformulated Schubert calculus. Additional \ physical refinements of these
ideas can be found in the work by Witten [33].

$H_{\ast}^{{}}$($G^{\mathbf{C}}(d,n);\mathbf{Z})$ is generated by the
\textit{special} Schubert cycles given by
\begin{equation}
\sigma(i)=\Omega(i,n-d+1,...,n) \tag{1.18}%
\end{equation}
for $i=0,1,...,n-d$. These results allow to prove [12,35,36] the following
theorem of central importance for the whole development presented in the rest
of this paper.

\bigskip

\textbf{Theorem 1.4.} For all sequences of integers 0 $\leq a_{0}%
<a_{1}<...<a_{d}\leq n$ (which we denote as \b{a} ) the following
determinantal (Giambelli's-like) formula holds in the homology ring $H_{\ast
}^{{}}$($G^{\mathbf{C}}(d,n);\mathbf{Z}):$%
\begin{equation}
\Omega(\text{\b{a}})=\left|  \sigma(a_{i}+j-i)\right|  \text{ , 0}\leq i,j\leq
n-d \tag{1.19}%
\end{equation}
with $\left|  \sigma(a_{i}+j-i)\right|  $ being a determinant made of special cycles.

\bigskip

\textbf{Remark 1.5}. The name Giambelli comes from the fact that structurally
the same expression exist for the Schur polynomials S$_{\text{\b{a}}}$\ which
was discovered by Giambelli [39-42]. The Schur polynomials are characters of
the general linear group on symmetrized complex linear vector space
$E^{\text{\b{a}}}$ [39-42]. In the light of results presented earlier we can
associate the determinant $p$($j_{0},...,j_{d})$ with that for special
Schubert cycles, Eq.(1.19), so that the Schur polynomials are tau functions of
KP hierarchy. The formal correspondence between the Schur polynomials and
Eq.(1.19) is not coincidental. It can be proven [42] that, actually, there is
an isomorphism $\Theta$ between S$_{\text{\b{a}}}$ and $\Omega($\b{a}$)$.

\textbf{Remark 1.6}. The topological meaning of Eq.(1.19) had been clarified
by Porteous [43] and also by Horrocks [44] and, later, by Carrel [45]. All
these results can be actually deduced directly from much earlier fundamental
papers by Chern [46] and Ehresmann [47].

We would like to describe briefly these results since they are essential for
correct physical understanding of the meaning of intersection numbers. To this
purpose let us observe that for $k$-cycle $A$ and $n-k$ cycle $B$ on an
$n$-dimensional manifold $M$ the \textit{Poincare'} \textit{duals} of these
cycles are closed $n-k$ and $k$ differential forms $\varphi$ and $\psi$
respectively so that the intersection number $^{\#}$($A\cdot B)$ of these
cycles in homology is equal to the wedge product of these two forms in
cohomology [12], i.e.
\begin{equation}
^{\#}(A\cdot B)=\int\limits_{M}\varphi\wedge\psi. \tag{1.20}%
\end{equation}
This definition can be extended to describe the intersection of subvarieties
$V$ and $W$ of a complex manifold $M$ and it is possible to prove that
Schubert calculus is just a special case of such more general algebra known in
the literature as \textit{Chow algebra} [8] (also as Chow ring [41]). Next, we
need a notion of a divisor.

If the manifold $M$ can be decomposed as
\begin{equation}
M=V_{1}\cup\cdot\cdot\cdot\cup V_{m} \tag{1.21}%
\end{equation}
then one introduces

\textbf{Definition 1.7}.Divisor $D$ on $M$ is locally finite formal linear
combination
\begin{equation}
D=\sum a_{i}V_{i} \tag{1.22}%
\end{equation}
of irreducible analytic hypersurfaces $V_{i}$ of $M$.

Definitions of ''local finitness'' and ''irreducibility'' are given in
Ref.[12] on page 130.The constants $a_{i\text{ }}$ are normally some integers.
For $G^{\mathbf{C}}(d,n)$ Schubert cycles provide the desired decomposition of
the Grassmannian [12]. The following theorem is of central importance [12].

\bigskip

\textbf{Theorem 1.8.}The Chern class\textbf{\ }$c_{1}(\lambda),$ given by
Eq.(1.9), of the line bundle $\lambda$\ represents the Poincare' dual of the
fundamental homology cycle carried by the divisor D, i.e.
\begin{equation}
\frac{i}{2\pi}\int\limits_{M}F\wedge\psi=\sum a_{i}\int\limits_{V_{i}}%
\psi\tag{1.23}%
\end{equation}
for every real closed $2n-2$ form $\psi.$

\bigskip

Consider now some implications of this theorem. First, in the case if $M$ is
compact Riemann surface, a divisor on $M$ is just a finite sum
\begin{equation}
D=\sum n_{i}p_{i} \tag{1.24}%
\end{equation}
of points $p_{i}\in M$ \ with multiplicities $n_{i}$. Combining Eqs.(1.23) and
(1.24) produces the Poincare'-Hopf index theorem
\begin{equation}
\frac{i}{2\pi}\int\limits_{M}F=\sum\limits_{i}n_{i}=\chi(M). \tag{1.25}%
\end{equation}
This theorem played central role in our earlier works of 2+1 gravity
[1,2].This observation clarifies both the meaning of numbers $n_{i}$ and
points $p_{i}$ : for vector (or line) fields these points are associated with
singularities of the field. These singularities had been interpreted as masses.

\textbf{Remark} \textbf{1.9}. Generalization of this result to higher
dimensions had been developed by Chern and Weil (CW). The up to date
exposition and generalization of their results can be found in Ref.[48] and,
in principle, provides an opportunity to describe 3+1 gravity in a way similar
to that developed for 2+1 dimensional case. Such approach to gravity is very
close in spirit to that originally suggested by Regge [49].

We need this observation in the present context as well for the following
reasons. By definition, a subset $V$ of an open set $U\in\mathbf{C}^{n}$ is an
\textit{analytic variety }if for any $p\in U$ there exists a neighborhood
$U^{^{\prime}}$ of $p$ in $U$ such that $V\cap U^{\prime}$ is \textit{common
zero locus} of a finite collection of holomorphic functions $\{f_{i}\} $ on
$U^{^{\prime}}.$ In particular, $V$ is called an analytic hypersurface if
locally $V$ is zero locus of a single nonzero holomorphic function $f$,
i.e.$V=\{f(z)=0\}$ in the neighborhood of $0\in V$. Accordingly, for
decomposition, Eq.(1.21), $V_{i}=\{f_{i}(z)=0\}$ with $V_{i} $ irreducible at
$0$ . All these facts lead to the following

\textbf{Definition 1.10}. An algebraic variety $V\subset\mathbf{P}^{n}$ is the
image in \textbf{P}$^{n}$ of \ zero locus of a collection of homogenous
polynomials defined in \textbf{C}$^{n+1}$, i.e. $V$=( $\mathcal{F}_{i}%
(z_{0},...,z_{n})=0).$

For the line bundle $\lambda$ on $M$ it is possible to associate sections with
$\mathcal{F}_{i}$ if $M$ can be embedded into \textbf{P}$^{n}($and, in our
case, it can be embedded since we had mentioned already that $\lambda=f^{\ast
}(\xi)$ with $\xi\in BU(n)$). Suppose now that for some point $z^{\ast}\in M$
of $n$-dimensional complex manifold $m$ sections are linearly dependent. Then,
according to CW theory [48,50], this fact can be written as
\begin{equation}
f_{\alpha_{1}}\wedge\cdot\cdot\cdot\wedge f_{\alpha_{m}}=0. \tag{1.26}%
\end{equation}
This equation is multidimensional analogue of the Poincare' condition (e.g.
read Remark 4.2. of Ref [2]) for the singularities of the line/vector fields
on surfaces. The \textit{degeneracy set} is set of all points $z_{i}^{\ast}$
for which condition given by Eq.(1.26) holds. In the most general case it is
$m-1$ dimensional submanifold \^{D} of $M$. Consider now a cycle $\alpha$ on
$M$ of dimension $r$ lesser than that of \^{D}. Let such a cycle meet \^{D}
transversely at the point $z_{i}^{\ast}\in$\^{D} then, $f$($\alpha)$ will meet
some Schubert cycle $\Omega($\b{a}$)$ also transversely at the point
$f(z_{i}^{\ast})$ of the Grassmannian so that the intersection number of
$f$($\alpha)$ with $\Omega($\b{a}$)$ at $f(z_{i}^{\ast})$ \ will be the same
as that for $\alpha$ meeting \^{D} on $M$. Taking into account Eq.s(1.20) and
(1.23) and also Theorem 1.8., we conclude that $c_{r}(\lambda)(\alpha)=$
$^{\#}(\alpha,$\^{D}$).$ With little additional work it can be shown [12]
that
\begin{equation}
\Omega(\alpha)=\left|  c_{\alpha_{i}+j-i}(\lambda)\right|  \text{ , }0\leq
i,j\leq n-d, \tag{1.27}%
\end{equation}
to be compared with Eq.(1.19). This result was obtained by Porteous [43] and
is known in the literature as Porteous formula [12,41].

\subsection{From Schubert varieties to directed random walks}

Porteous formula can be seen as a special case of much more comprehensive
result of Weil (and developed by Chern ) known as \textit{Weil homomorphism}.
In view of this homomorphism, the result given by Eq.(1.27) can be
reinterpreted as Schur polynomial (e.g. read discussion following Eq.(1.19))
and, hence, as $\tau$ function of KP hierarchy. Schur polynomials S$_{\lambda
}$ originate from the known [40] identity for indeterminates $\{x_{i}\}$%
\begin{equation}
\left(  x_{1}+\cdot\cdot\cdot+x_{m}\right)  ^{n}=\sum\limits_{\lambda\vdash
n}f^{\lambda}\text{S}_{\lambda}(x_{1},...,x_{m}), \tag{1.28}%
\end{equation}
where use of the notation $\lambda\vdash n$ is meant to say that $\lambda$ is
partition of $n$. Partitions are best represented by the Young tableaux.
Accordingly, the factor $f^{\lambda}$ denotes a number of standard tableaux of
a given shape $\lambda$ .

Eq.(1.28) can be rewritten in a slightly different form as follows
\begin{equation}
\left(  x_{1}+\cdot\cdot\cdot+x_{n}\right)  ^{M}=\sum\limits_{\left(
m_{1},...,m_{n}\right)  }\frac{M!}{m_{1}!m_{2}!...m_{n}!}x_{1}^{m_{1}}%
\cdot\cdot\cdot x_{n}^{m_{n}} \tag{1.29}%
\end{equation}
provided that $M=m_{1}+\cdot\cdot\cdot+m_{n}.$ This observation allows us to
write Schur polynomial S$_{\lambda}$ in the form
\begin{equation}
\text{S}_{\lambda}(x_{1},...,x_{n})=\sum\limits_{m=\left(  m_{1}%
,...,m_{n}\right)  }K_{\lambda,m}x_{1}^{m_{1}}\cdot\cdot\cdot x_{n}^{m_{n}}
\tag{1.30}%
\end{equation}
with $K_{\lambda,m}$ being some coefficients (Kostka numbers) [42].

Eq.(1.30) can be given probabilistic meaning in terms of the directed random
walks. Indeed, following Ref.[42,51], consider planar lattice. On this lattice
consider a directed path P from (a,1) to (b,N). The information about this
path can be encoded into multiset $Hor_{y}$(P) of $y-$coordinates of the
horizontal steps of P. Define
\begin{equation}
\text{w}(\text{P})=\prod\limits_{i=Hor_{y}(\text{P})}x_{i} \tag{1.31}%
\end{equation}
To facilitate reader's understanding, we illustrate these ideas on Fig.2.
\begin{figure}
[ptb]
\begin{center}
\includegraphics[
natheight=4.478900in,
natwidth=6.156600in,
height=2.3938in,
width=3.2802in
]%
{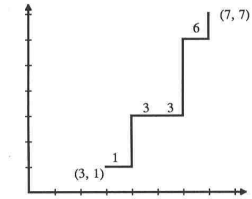}%
\caption{A typical directed random walk}%
\end{center}
\end{figure}
In this figure $Hor_{y}$(P)=\{1,3,3,6\} and, accordingly, w(P)=$x_{1}^{1}%
x_{3}^{2}x_{6}^{1}$. Next, we need to extend this result to an assembly of
directed random walks (''vicious'' random walkers in terminology of Fisher
[52]). That is we need to consider the products of the type w(P$_{1}%
)\cdot\cdot\cdot$w(P$_{k})\equiv$W$($\^{P}). Finally, the generating function
for an assembly of such vicious walkers is given by
\begin{equation}
h_{b-a}(x_{1},...,x_{N})=\sum\limits_{\text{\^{P}}}\text{W}(\text{\^{P}})
\tag{1.32}%
\end{equation}
where W(\^{P}) is made of monomials of the type $x_{1}^{m_{1}}x_{2}^{m_{2}%
}...x_{N}^{m_{N}}$ provided that $m_{1}+\cdot\cdot\cdot+m_{N}=b-a.$ The
following theorem is proven in Ref.[42,51]

\bigskip

\textbf{Theorem 1.11}. Given integers 0 $<a_{1}<...<a_{k}$ and $0$
$<b_{1}...<b_{k}$ , let M$_{i,j}$ be the $k\times k$ matrix
\begin{equation}
\text{M}_{i,j}=h_{b_{j}-a_{i}}(x_{1},...,x_{N}), \tag{1.33}%
\end{equation}
then,
\begin{equation}
\det\text{M=}\sum\limits_{\text{\^{P}}}\text{W}(\text{\^{P}}), \tag{1.34}%
\end{equation}
where the sum is taken over all sequences (P$_{1},...,$P$_{k})=$\^{P} of
non-intersecting lattice paths P$_{i}:$ ($a_{i},1)\rightarrow(b_{i},N).$

\bigskip

\textbf{Corollary 1.12.} Put now $a_{i}=i$ and $b_{j}=\lambda_{i}+j$ in
Eq.(1.33) provided that \ $1\leq i,j\leq k$ with $\lambda$ being partition of
$N$ with $k$ parts then, $\det$M=S$_{\lambda}(x_{1},...,x_{N}).$

Theorem 1.11. provides desired connection between the vicious random walkers
and the $\tau$ function of KP hierarchy.

\subsection{Organization of the rest of this paper}

In Section 2 we provide some facts from the theory of random vicious walkers
using results of Fisher [52], Huse and Fisher [53] and Forrester [54].We argue
that these results can be obtained also with help of the Bethe ansatz method
applied to one dimensional non ideal Bose gas. Such observation is helpful for
developing connections between the Yang-Baxter equation, symmetric functions
and Schubert polynomials. The obtained Bethe ansatz wave function is
reinterpreted in terms of the Gaussian unitary ensemble of random matrices.
Using some results for random matrices summarized by Mehta [55] and more
recent results by Tracy and Widom [56] (and Forrester [57]) we discuss
relations between the Kontsevich matrix Airy integral and that coming from the
Gaussian unitary ensemble. We argue that both integrals are tau functions of
KP hierarchy. Moreover, we demonstrate that these integrals are actually tau
functions for KdV hierarchy of equations (that is they are only a special case
of KP hierarchy) and, hence, both can be used as solutions of the W-K model.
The arguments of Section 2 are too general. They do not contain explicit
reference to 2+1 gravity, moduli space, etc. This deficiency is corrected in
Sections 3 and 4 which contain new combinatorial proof of the main identity
obtained by Kontsevich [9]:
\begin{equation}
\sum\limits_{\Sigma d_{i}=3g-3+n}<\tau_{d_{1}}\cdot\cdot\cdot\tau_{d_{n}}%
>_{g}\prod\limits_{i=1}^{n}\frac{(2d_{i}-1)!!}{\lambda_{i}^{2d_{i}+1}}%
=\sum\limits_{G\in\Gamma_{g,n}}\frac{2^{-\mathcal{V}(\Gamma)}2^{\mathcal{E}%
(\Gamma)}}{\left|  Aut(G)\right|  }\prod\limits_{e\in\mathcal{E}(G)}\frac
{1}{\lambda_{i(e)}+\lambda_{j(e)}}. \tag{1.35}%
\end{equation}
This identity connects the observables of topological quantum gravity, Eq.
(1.2), with averages of the random matrix (Kontsevich) model associated
effectively with the ribbon graphs $\Gamma_{g,n}$ representing the
combinatorial moduli space $\mathcal{M}_{g,n}^{comb}$ of marked Riemann
surfaces of genus $g$. To avoid repeats of technical details describing random
matrices, ribbon graphs, etc., discussed in Sections 3 and 4, we only notice
that $\mathcal{V}$($\Gamma)$ in Eq.(1.35) represents the total number of
vertices and $\mathcal{E}$($\Gamma)$ the total number of edges of the ribbon
graph $\Gamma_{g,n}.$ Mathematically, the identity, Eq.(1.35), is just the
statement that there are different but equivalent ways to present partitions
discussed in Section 1.4. In section 3 we connect these partitions with some
results coming from Nielsen-Thurston theory of surface automorphisms. As it
was demonstrated in Refs.[1-4], this theory provides natural mathematical
framework for description of dynamics of 2+1 gravity. In these references the
canonical (fixed genus $g$) partition function for 2+1 gravity was obtained.
This partition function was used for description of the dynamical transition
from the pseudo-Anosov to Seifert-fibered (periodic) regime (phase) of 2+1
gravity. We remind our readers about these concepts within the boundaries of
Nielsen-Thurston theory. In Sections 3 and 4 we argue that the W-K partition
function is relevant for the Seifert-fibered phase. This phase was not
discussed in Refs.[1-4]. We should warn our readers that complete description
of the Sefert-fibered phase requires more than just the W-K model. For
instance, Kulkarni and Raymond [58] had found a very interesting connection
between the Seifert-fibered \ and anti-de Sitter 3 manifolds which are just
Lorentz manifolds of constant negative curvature. Such anti-de Sitter 3
manifolds had been recently classified by Francois [59]. In spite of the fact
that they have received considerable attention in physics literature [3]
recently, the complete description of this phase ( which by the way contains
all known crystallographic groups and much more [60] ) in physical terms
remains challenging research problem.

Thus,the results of Refs.[1,4] for the partition function of 2+1 gravity
should be extended to reach an accord with results of W-K model. To this
purpose, following Nielsen [61], dynamics of the Riemann surface
homeomorphisms should be lifted to the universal cover, e.g. to the Poincare'
disc model of \textbf{H}$^{2}$. In such model the set of geodesics (geodesic
lamination) is represented by the set of non intersecting arcs (i.e. circular
segments whose ends lie on $S_{\infty}^{1}).$The combinatorial arrangement on
the disc is described by the Catalan numbers $C_{n}$. These numbers had been
used earlier in our work, Ref.[1] for construction of the partition function.
In this paper we use this combinatorial data in order to establish several
important bijections:1) from the set of arcs to the set of Dyck paths, Fig.5;
2) from the set of Dyck paths to the set of parallelogram polyominoes, Fig.6;
3) from the set of polyominoes to the set of Young tableaux and, finally, 4)
from the set of Young tableaux to the set of vicious walkers, Figs.7 and 8.
This chain of bijections is needed in order to bring our earlier obtained
results for the partition function in accord with the l.h.s. of the Kontsevich
identity, Eq.(1.35).To bring this partition function in accord with the r.h.s.
of this identity the ribbon graphs need to be constructed. This is discussed
in Section 4. Unlike Kontsevich [9] and others [62-64 ], we do not use
quadratic differentials (discussed earlier in our papers, Refs.[1,2] ) for
construction of these graphs. Our method is based on use of known various
equivalent geometrical ways to describe the combinatorics of Catalan numbers
[42]. The edges of thus constructed ribbon graphs are replaced by the paths of
vicious walkers so that in the end using combinatorics of the Young tableaux
the partition function for assembly of vicious walkers acquires the form of
the r.h.s. of Eq.(1.35).

Section 5 is included in this work for several reasons. First, it discusses
the issue of universality of the W-K model from the point of view of dynamical
systems theory. By universality we mean the fact that partition functions of
W-K and many related models to be discussed briefly in this section are
solutions of the KdV hierarchy. The reasons for this universality can be
traced back to the very basic properties of quasiconformal transformations
known already to Ahlfors long time ago [65]. Although in Ref. [3] these
transformations had been discussed extensively, in this paper more up to date
information is presented to make it relevant to the results obtained in
earlier sections. This includes some facts about the Thompson and the Ptolemy
groups, about their connections with binary trees (and, hence, with Catalan
numbers) and their relations with the universal Teichm\"{u}ller and moduli
spaces, etc. Second, it discusses briefly some potential physical and
biological applications of the obtained results. Finally, it discusses
connections with Frobenius manifolds, self-dual Einstein equations, etc., thus
leaving many problems open for further study.

\section{ From vicious walkers to Kontsevich model via \ Gaussian unitary
ensemble of random matrices}

Following Forrester [54] and Fisher [52], we would like now to formulate a
lock step model of vicious walkers. Incidentally, this model is just many
walks generalization of the directed polymer model considered in our earlier
work, Ref.[66].The continuum limit of the distribution function for this
directed polymer produces \ the Euclidean version of the Dirac propagator for
particle whose mass is associated with bending probabilities to be discussed
below. In case of many walkers the fermionic nature of the Dirac ''particles''
(walks) imposes a sort of Pauli principle which forbids two walks to
intersect. This is characteristic to all quantum many body problems where all
''particles'' live in the same ''world time'' [67]. In the theory of Brownian
motion each walker is allowed to have its own world time [68] so that,
accordingly, one can have many \ world times quantum mechanics. We shall
refrain from discussion of these options referring interested reader to
current literature [69]. The lock step model assumes just one ''world time''.
Technically this means the following.

We consider a square lattice where $x$ coordinate is assigned for ''space''
while the $y$ coordinate is assigned for ''time''. If $p$ walkers are labelled
in linear sequence along $x$ axis so that one has
\begin{equation}
x_{1}<x_{2}\cdot\cdot\cdot<x_{p} \tag{2.1}%
\end{equation}
at ''time'' $t=0$, the same inequalities should hold for all subsequent times.
The walkers start either on the even or on the odd numbered sites on the x
axis. At each tick of the clock each walker moves either to the right or to
the left (along lattice diagonals) with equal (bending) probability $w$. The
probabilities, in general, may not be equal and are associated with masses of
the Dirac particles as discussed in our work, Ref.[66]. Because of imposed
initial condition, no two walkers can occupy the same lattice space at any
time. This circumstance makes walkers ''vicious''. Let \textbf{x}%
$_{0}=(x_{1,0,...,}x_{p,0})$ be the initial configuration of such vicious
walkers and \textbf{x}=($x_{1},...,x_{p})$ be the final configuration at time
$t$. To calculate the total number of walks starting at $t=0$ at
\textbf{x}$_{0}$ and ending at $t$at \textbf{x} we need to know the
probability distribution $W_{p}(\mathbf{x}_{0}\rightarrow\mathbf{x};t)$ that
the walkers proceed without passing, i.e. maintaining the inequalities
\begin{equation}
x_{j-1}(t^{\prime})<x_{j}(t^{\prime})\text{ , }j=2,3,...,p\text{\ for }0\leq
t^{\prime}\leq t \tag{2.2}%
\end{equation}
from their initial configuration\ at time $t=0$ to their final configuration
at time $t$. It is of interest to study this problem for large times:
$t\rightarrow\infty$ . In this case it is possible to make a transition to the
continuum limit in our study of the probability distribution. In doing so, we
shall follow the arguments of Huse and Fisher [53]. To begin, we need to
recall that in the continuum limit the probability distribution $W_{p}%
^{0}(\mathbf{x}_{0}\rightarrow\mathbf{x};t)$ for $p$ \textbf{independent}
random walkers is known to be
\begin{equation}
W_{p}^{0}(\mathbf{x}_{0}\rightarrow\mathbf{x};t)=\exp\{-\frac{\left|
\mathbf{x}-\mathbf{x}_{0}\right|  ^{2}}{2Dt}\}/(2\pi Dt)^{p/2}. \tag{2.3}%
\end{equation}
The diffusion constant $D$ sets up the scale since, as usual,
$<$%
(x$_{j}-x_{j,0})^{2}>=Dt.$ Because of \ use of a single world time for all
walkers the above distribution function can be also viewed as distribution
function for a \textbf{single} walker in $p$-dimensional Euclidean space. The
restrictions \ given by Eq.(2.2) impose additional constraints that such a
walker must not cross any (hyper)planes described by the set of equations
$x_{1}=x_{2},x_{2}=x_{3},...,x_{p-1}=x_{p}$. In view of the results of Section
1, e.g. see Fig.2 and discussion related to it, the assembly of these planes
forms Grassmann manifold. This statement can be clarified further by
considering the following example. Following Gaudin [70] consider
Schr\"{o}dinger equation for one dimensional Bose gas\ of $p$ particles
interacting via point-like pairvise interaction potential. The dimensionless
form of the Schr\"{o}dinger equation for such particle system is given by
\begin{equation}
-\sum\limits_{i=1}^{p}\frac{\partial_{i}\Psi}{\partial_{i}x^{2}}%
+2c\sum\limits_{i<j}\delta(x_{i}-x_{j})\Psi=E\Psi. \tag{2.4}%
\end{equation}
This equation is equivalent to the boundary value problem of obtaining the
wave function $\Psi$ of the equation
\begin{equation}
-\Delta_{p}\Psi=E\Psi, \tag{2.5}%
\end{equation}
where $\Delta_{p}$ is just $p$-dimensional ''free'' Laplacian, and the wave
function $\Psi$ is subjected to the set of constraints:
\begin{equation}
\left(  \frac{\partial}{\partial x_{i}}-\frac{\partial}{\partial x_{j}%
}\right)  \Psi\mid_{x_{i}-x_{j}=0^{+}}=2c\Psi,1\leq i<j\leq p. \tag{2.6}%
\end{equation}
The condition $\Psi$=0 on the hyperplanes is achieved in the limit $c$%
=$\infty$ according to Gaudin [70]. It can be shown, that for any $c$ solution
of the corresponding quantum mechanical problem is obtained with help of the
Bethe ansatz method [70]. Moreover, any problem solvable by the Bethe ansatz
method is essentially of the type just described as had been demonstrated
rigorously by Gutkin [71]. Therefore, not surprisingly, that there are deep
connections between the classical exactly integrable systems of KP type and
the quantum mechanical exactly integrable systems solved by the Bethe ansatz
method [72]. The Hecke algebra leading to the Yang-Baxter equations providing
mathematical justification of the Bethe ansatz method is coming from some
particular representation of the symmetric group [73] and, hence, is connected
with Schur and related polynomials. Some additional details can be found in
Ref. [74].

The mathematical problem posed by Eq.s(2.5) and (2.6) can be equivalently
formulated as problem about properties of random walk inside $p-$dimensional
caleidoscope. That is we are looking for solution of an eigenvalue problem for
''free '' Laplacian compatible with some reflection group[70,71]. In the
simplest cases these are just subgroups of the symmetric group $S_{p}$ made
out of even and odd permutations. In view of this, we obtain,
\begin{equation}
W_{p}(\mathbf{x}_{0}\rightarrow\mathbf{x};t)=\sum\limits_{g\in S_{p}%
}\varepsilon(g)W_{p}^{0}(g\mathbf{x}_{0}\rightarrow\mathbf{x};t), \tag{2.7}%
\end{equation}
where $\varepsilon(g)=\pm1$ depending upon the symmetry of permutation (even
or odd). Taking Eq.(2.3) into account, this result can be rewritten
equivalently as
\begin{equation}
W_{p}(\mathbf{x}_{0}\rightarrow\mathbf{x};t)=U_{p}(\mathbf{x}_{0}%
,\mathbf{x};t)\frac{\exp\{-(\mathbf{x}^{2}+\mathbf{x}_{0}^{2})/2Dt\}}{\left(
2\pi Dt\right)  ^{p/2}}, \tag{2.8}%
\end{equation}
where
\begin{equation}
U_{p}(\mathbf{x}_{0},\mathbf{x};t)=\sum\limits_{g\in S_{p}}\varepsilon
(g)\exp\left[  (\mathbf{x\cdot}g\mathbf{x}_{0})/Dt\right]  . \tag{2.9}%
\end{equation}
Some short calculation explained in Ref. [53] produces
\begin{equation}
U_{p}(\mathbf{x}_{0},\mathbf{x};t)\simeq const\Delta(\mathbf{x})\Delta
(\mathbf{x}_{0})/\left(  Dt\right)  ^{n_{p}} \tag{2.10}%
\end{equation}
with $const=1/1!2!...(p-1)!$ , $n_{p}=\frac{1}{2}p(p-1)$ and $\Delta
(\mathbf{x})$ being the Vandermonde determinant:
\begin{equation}
\Delta(\mathbf{x})=\prod\limits_{i<j}(x_{i}-x_{j}). \tag{2.11}%
\end{equation}
\textbf{Remark 2.1.} \ From standard texts in probability theory, e.g. see
Ref.[75], it is known that non normalized expression for the probability
$W_{p}^{0}(\mathbf{x}_{0}\rightarrow\mathbf{x};t)$ is the long time limit of
the formula providing the total number of walks of $n$ steps (since
$t\rightleftharpoons n)$ from point \textbf{x}$_{0}$ \ to point \textbf{x}.
Accordingly, Eq.s (2.7)-(2.10) provide the total number of nonintersecting
directed walks and, hence, $W_{p}(\mathbf{x}_{0}\rightarrow\mathbf{x}%
;t)\approx\det$ M as shown in Eq.(1.34).

It is convenient to assume now that \textbf{x}=\textbf{x}$_{0}.$ Then, upon
rescaling, the following result holds:
\begin{equation}
W_{p}(\mathbf{x=x}_{0};t)\equiv P_{p}(x_{1},...,x_{p})=const\times
\exp(-\mathbf{x}^{2})\Delta^{2}(\mathbf{x}). \tag{2.12}%
\end{equation}
This is just the probability distribution of eigenvalues of random matrices
from the Gaussian unitary ensembe [55]. Following Dyson [76] \ define the
$n-$point correlation function by
\begin{equation}
R_{n}(x_{1},...,x_{n})=\frac{p!}{\left(  p-n\right)  !}\int\limits_{-\infty
}^{\infty}\cdot\cdot\cdot\int\limits_{-\infty}^{\infty}P_{p}(x_{1}%
,...,x_{p})dx_{n+1}\cdot\cdot\cdot dx_{p}. \tag{2.13}%
\end{equation}
Using this definition the connected $n$-point correlation functions
$T_{n}(x_{1},...,x_{n})$ are defined in a usual fashion, e.g.
\[
T_{1}(x)=R_{1}(x)
\]%
\[
T_{2}(x_{1},x_{2})=-R_{2}(x_{1},x_{2})+R_{1}(x_{1})R_{2}(x_{2}),
\]
etc. Using method of orthogonal polynomials [55] it can be demonstrated that
\begin{equation}
R_{n}=\det[K_{p}(x_{i},x_{j})]_{\mid i,j=1,...,n} \tag{2.14}%
\end{equation}
where the kernel $K_{p}(x_{i},x_{j})$ is given by
\begin{equation}
K_{p}(x,y)=\sum\limits_{k=0}^{p-1}\varphi_{k}(x)\varphi_{k}(y) \tag{2.15}%
\end{equation}
with functions $\varphi_{k}(x)$ depending on the random matrix ensemble used.
In the case of Gaussian unitary ensemble they are given by
\begin{equation}
\varphi_{k}(x)=(2^{k}k!\sqrt{\pi})^{-\frac{1}{2}}\exp(x^{2}/2)(-\frac{d}%
{dx})^{j}\exp(-x^{2}). \tag{2.16}%
\end{equation}
The connected $n$-point correlation function $T_{n}(x_{1},...,x_{n})$ can be
neatly represented using the kernel $K_{p}(x,y)$ as follows [55], page 92,:
\begin{equation}
T_{n}(x_{1},...,x_{n})=\sum\limits_{P}K_{p}(x_{1},x_{2})K_{p}(x_{2}%
,x_{3})\cdot\cdot\cdot K_{p}(x_{n},x_{1}), \tag{2.17}%
\end{equation}
where the sum is over all $(n-1)!$ distinct cyclic permutations of indices
$(1,...,n)$. For $p\rightarrow\infty$ the kernel $K_{p}(x,y)$ can be
calculated and in terms of the rescaled variables it was obtained
independently by Tracy and Widom [56] and by Forrester [57]:
\begin{equation}
K(X,Y)=\frac{Ai(X)Ai^{\prime}(Y)-Ai(Y)Ai^{\prime}(X)}{X-Y}, \tag{2.18}%
\end{equation}
where $Ai(y)$ denotes the Airy function
\begin{equation}
Ai(y)=\int\limits_{-\infty}^{\infty}\exp(i(x^{3}/3-xy))dx \tag{2.19}%
\end{equation}
and the prime denotes differentiation with respect to its argument. Following
Kontsevich [9] we define now the matrix Airy function analogous to the
''scalar'' case
\begin{equation}
A(\mathbf{Y})=\int\exp(i\text{ }tr(\mathbf{X}^{3}/3-\mathbf{XY}))\mathbf{dX}
\tag{2.20}%
\end{equation}
where $\mathbf{X}$ and $\mathbf{Y}$ are hermitian $N\times N$ matrices for
some $N.$ After some computation Kontsevich obtains,
\begin{equation}
A(\mathbf{Y})=\left(  2\pi\right)  ^{N(N-2)/2}\frac{\det(A^{(j-1)}(Y_{i}%
))}{\det(Y_{i}^{j-1})}, \tag{2.21}%
\end{equation}
where
\begin{equation}
A^{(j-1)}(y_{i})=\int\limits_{-\infty}^{\infty}dxx^{j-1}\exp(i(x^{3}%
/3-xy_{i}))=\left(  i\frac{\partial}{\partial y_{i}}\right)  ^{j-1}Ai(y_{i}).
\tag{2.22}%
\end{equation}
For $i,j=1-2$ we obtain,
\begin{equation}
A(X,Y)/(2\pi)^{1/2}=K(X,Y), \tag{2.23}%
\end{equation}
that is the Tracy-Widom kernel and the Kontsevich Airy matrix integral are
practically identical. Naturally, it is of interest to find out if this result
will hold for $i,j>2$. Comparing Eq.s. (2.14),(2.17) and (2.23) we conclude,
that Eq.(2.17) should be considered as a likely candidate for further
treatment. This conclusion is in accord with recent results of Okounkov [77].
We shall use some of his results below in Section 3 while in this section we
would like to discuss different approach. To this purpose, following Mehta
[55] , let us notice that correlation function $R_{n}$ , Eq.(2.13), can be
presented in the following form
\begin{equation}
R_{n}=\sum\limits_{P}(-1)^{n-m}\prod\limits_{1}^{m}K_{p}(x_{a},x_{b}%
)K_{p}(x_{b},x_{c})\cdot\cdot\cdot K_{p}(x_{d},x_{a}) \tag{2.24}%
\end{equation}
where the permutation $P$ is a product of $m$ exclusive cycles of lengths
$h_{1},h_{2},...,h_{m}$ of the form ($a\rightarrow b\rightarrow c\rightarrow
\cdot\cdot\cdot\rightarrow d\rightarrow a),\sum\nolimits_{1}^{m}h_{j}=n.$
Comparison between Eqs.(2.17) and (2.24) indicates that if, say, the connected
$n-$point correlation function $T_{n}$ is tau function of KP hierarchy, then
correlation function $R_{n}$ should possess this property as well.

To prove that both $R_{n\text{ }}$and $\ T_{n}$ are indeed tau functions
several steps are required. First, we would like to reconsider Eq.(2.21) in
the light of subsequent refinements of Kontsevich work in physics literature.
Following Di Francesco [78] the Kontsevich integral $\Theta_{N}(\Lambda)$ is
given by
\begin{equation}
\Theta_{N}(\Lambda)=\frac{\int e^{tr(i\frac{M^{3}}{6}-\frac{\Lambda M^{2}}%
{2})}dM}{\int e^{-tr\frac{\Lambda M^{2}}{2}}dM}, \tag{2.25}%
\end{equation}
where $\Lambda$ is diagonal $N\times N$ real matrix with elements
$\lambda=(\lambda_{1},...,\lambda_{N})$ along the diagonal and $M$ being
$N\times N$ Hermitian matrix. After some calculations this integral can be
brought to the following form:
\begin{equation}
\Theta_{N}(\Lambda)=\frac{\left|  z,Dz,D^{2}z,...,D^{N-1}z\right|  }{\left|
1,\lambda,\lambda^{2},...,\lambda^{N-1}\right|  }, \tag{2.26}%
\end{equation}
where
\begin{equation}
z=z(\lambda)=\int\limits_{-\infty}^{\infty}dm\sqrt{\frac{\lambda}{2\pi}}%
\exp\left[  i(\frac{m^{3}}{6}+\frac{m\lambda^{2}}{2}-\frac{\lambda^{3}}%
{3})\right]  , \tag{2.27}%
\end{equation}
and
\begin{equation}
D=\lambda+\frac{1}{2\lambda^{2}}-\frac{1}{\lambda}\frac{d}{d\lambda}.
\tag{2.28}%
\end{equation}
The Vandermonde determinant $\Delta(\lambda)$, e.g. see Eq.(2.11), is written
in the present case as
\begin{equation}
\Delta(\lambda)=\left|  1,\lambda,\lambda^{2},...,\lambda^{N-1}\right|
\tag{2.29}%
\end{equation}
so that the expression in the numerator of Eq.(2.26) is also a determinant.
Second, since $z(\lambda)$ is a solution of Airy's equation
\begin{equation}
\left(  D^{2}-\lambda^{2}\right)  z(\lambda)=0 \tag{2.30}%
\end{equation}
written in a somewhat unconventional form, it is possible to replace terms of
the type $D^{2p}z(\lambda)$ in the determinant of Eq.(2.26) by $\lambda
^{2p}z(\lambda)$ and, analogously, $D^{2p+1}z(\lambda)$ by $\lambda^{2p+1}%
\bar{z}(\lambda)$ where
\begin{equation}
\bar{z}(\lambda)=\frac{1}{\lambda}Dz(\lambda). \tag{2.31}%
\end{equation}
This allows us to rewrite $\Theta_{N}(\Lambda)$ in the following form
\begin{equation}
\Theta_{N}(\Lambda)=\frac{\left|  x^{N-1}z,x^{N-2}\bar{z},....\right|
}{\left|  x^{N-1},x^{N-2},...,1\right|  } \tag{2.32}%
\end{equation}
with $x=1/\lambda.$ Third, if the asymptotic expansions of $z$ and $\bar{z}$
given by
\begin{equation}
z(\lambda)=\sum\limits_{k\geq0}c_{k}\lambda^{-3k} \tag{2.33}%
\end{equation}%
\begin{equation}
\bar{z}(\lambda)=\sum\limits_{k\geq0}d_{k}\lambda^{-3k} \tag{2.34}%
\end{equation}
with known coefficients $c_{k}$ and $d_{k}$ are substituted into Eq.(2.32) it
acquires the following final form:
\begin{equation}
\Theta_{N}(\Lambda)=\sum\limits_{n_{1},...,n_{N}\geq0}\prod\limits_{i=0}%
^{N}a_{n_{i}}^{\left(  i\operatorname{mod}2\right)  }\frac{\left|
x^{3n_{1}+N-1}z,x^{3n_{2}+N-2}\bar{z},....,x^{3n_{N}}\right|  }{\left|
x^{N-1},x^{N-2},...,1\right|  }. \tag{2.35}%
\end{equation}
>From this form one can recognize at once the Jacobi-Trudy formula [39-42] for
the Schur polynomials S$_{n}$ given as the ratio of determinants. Since Schur
polynomials are $\tau$ functions of the KP hierarchy as we had discussed in
Section 1, it is clear that $\Theta_{N}(\Lambda)$ is also $\tau$ function of
KP hierarchy. Due to specific form of this function (it contains only the odd
powers of $\lambda)$ such tau function is actually tau function for the KdV
hierarchy in accord with Miwa et all [34]. It remains to demonstrate now that
Eq.s(2.17) and (2.24) also can serve as tau functions of KdV hierarchy.
Evidently, for $n=2$ this is the case in view of the arguments just presented.
To prove that this is the case for $n>2$ it is sufficient to employ the
Littlewood-Richardson (''fusion'' formula in physics terminology) rule given
by
\begin{equation}
\text{S}_{\mu}\cdot\text{S}_{\nu}=\sum\limits_{\lambda}C_{\mu\nu}^{\lambda
}\text{S}_{\lambda} \tag{2.36}%
\end{equation}
with the Littlewood-Richardson coefficients $C_{\mu\nu}^{\lambda}$ assumed to
be known [39-42] in principle. Successive applications of this formula to
Eq.s(2.17) and (2.24) produces a combination of Schur polynomials each of
which is tau function of KP hierarchy. Moreover, since for $n=2$ such tau
function was that for KdV hierarchy, evidently, for the same reasons as
discussed by Di Francesco [78] the general case $n>2$ also produces tau
functions for KdV hierarchy.

\section{Nielsen-Thurston surface automorphisms and partition function of 2+1 gravity}

\subsection{Review of Nielsen-Thurston theory}

Let $\mathcal{S}$ be closed orientable Riemann surface of genus $g$.The first
homotopy group, the fundamental group $\pi_{1}(\mathcal{S})$ of surface
$\mathcal{S}$ is made of $2g$ generators $\{x_{i},y_{i}\}$, $i=1-g$ and a
single relation so that its presentation is known to be
\begin{equation}
\pi_{1}(\mathcal{S})=<x_{1},y_{1},...,x_{g},y_{g}\mid\lbrack x_{1,}y_{1}%
]\cdot\cdot\cdot\lbrack x_{g},y_{g}]>. \tag{3.1}%
\end{equation}
Nielsen has noted that there is one to one correspondence between
automorphisms of $\pi_{1}(\mathcal{S})$ and surface self-homeomorphisms. This
is summarized in the following proposition

\bigskip

\textbf{Proposition 3.1}.(Nielsen [68]) If $g>1$, then every element of
$Out$($\pi_{1}(\mathcal{S}))$ is represented by a unique isotopy class of
self-homeomorphisms of $\mathcal{S}$.

\bigskip

An important subgroup of $Out$($\pi_{1}(\mathcal{S}))$ is the mapping class
group $\mathcal{M}_{g}$ discussed in Section 1. Geometrically, this group is
finitely generated by the \textit{Dehn twists} \ in simple closed curves
(lamination set) on $\mathcal{S}$ whose physical significance we had discussed
extensively in our previous work, Ref.[4].A simple closed curve $\mathcal{C}$
on an orientable surface $\mathcal{S}$ has a neighborhood $\mathcal{E}$
homeomorphic to an annulus which is convenient to parametrize by
\{[$r$,$\theta$]$\mid$1$\leq r<2\}$.The Dehn twist in $\mathcal{C}$ can be
imagined as an automorphism T$_{C}:\mathcal{S}\rightarrow\mathcal{S}$. It is
given by the identity off $\mathcal{E}$ and by [$r$,$\theta$]$\rightarrow
\lbrack r$,$\theta+2\pi r]$ on $\mathcal{E}$ . Using results of our previous
works [1-4], let us illustrate these concepts on the simplest example of a
punctured torus T$^{2}$. In this case $Out$($\pi_{1}($T$^{2}))=GL_{2}%
(\mathbf{Z})$ and $\mathcal{M}_{1,1}=PSL(2,\mathbf{Z}).$ Since any
transformation from $PSL(2,\mathbf{Z})$ is obtainable by projectivisation of
$SL(2,\mathbf{Z})$ we discuss everything in terms of $SL(2,\mathbf{Z})$ with
projectivisation at the end. Any transformation which belongs to
$SL(2,\mathbf{Z})$ is expressible in terms of $2\times2$ \ matrix \textbf{A}
given by
\begin{equation}
\mathbf{A}=\left(
\begin{array}
[c]{cc}%
a & b\\
c & d
\end{array}
\right)  \tag{3.2}%
\end{equation}
with integer coefficients subject to condition:$\det$\textbf{A}=$ab-cd$=1. The
characteristic polynomial for this matrix is given by
\begin{equation}
t^{2}-tr\mathbf{A}\text{ }t+\det\mathbf{A}=0. \tag{3.3}%
\end{equation}
This implies that the eigenvalues of \textbf{A} are either :

a) both complex ( when tr\textbf{A}=0,1,-1),

b) both equal to $\pm1$ (when tr\textbf{A}=$\pm2),$

c) distinct and real (when $\left|  tr\mathbf{A}\right|  $
$>$%
2). \ \ \ 

If $\frak{F}_{A}$ is toral automorphism then, transformation a) is called
\textit{periodic} since $\left(  \frak{F}_{A}\right)  ^{n}=1$ for some $n$
(actually, $n=12$ by the Hamilton-Cayley theorem) , transformation b) is
called \textit{reducible} since it leaves a simple closed curve $\mathcal{C}$
invariant, transformation c) is called (\textit{pseudo) Anosov ((pseudo)}%
Anosov if the line/vector field on surface (does) does not contain
singularities). Such transformations are of an infinite order. Physical
significance of this fact is explained and illustrated in our previous work,
Ref.[4]. The largest of two eigenvalues is associated with topological entropy
of the line/vector flow and is related to the amount of stretching of surface
$\mathcal{S}$ and, hence, with the dilatation parameter of the Teichm\"{u}ller theory.

Nielsen-Thurston theory generalizes the above classification of surface
automorphisms to all surfaces of genus $g>1$.Already Nielsen had realized [61]
that for $g>1$ it is more convenient to study homeomorphisms of surface
$\mathcal{S}$ by considering their image on the universal cover of
$\mathcal{S}$ which we choose as Poincare' disc model of \textbf{H}$^{2}$,
i.e. $int$\textbf{D} $\cup$ $S_{\infty}^{1}=$\textbf{H}$^{2}.$According to
Nielsen [61]

\bigskip

\textbf{Proposition 3.2}. Any lift \~{h} of \ the surface self- homeomorphism
h: $\mathcal{S}\rightarrow\mathcal{S}$ to the universal cover of $\mathcal{S}$
extends to a unique self-homeomorphism of the unit disc \textbf{D}, i.e. to
int\textbf{D}$\cup S_{\infty}^{1}.$

\bigskip

If surface self-homeomorphisms h are associated with Dehn twists connected
with a set of simple closed nonintersecting curves homotopic to geodesics
(such set is called geodesic lamination $\mathcal{L}$), then their lifts
\~{h}($\mathcal{L}$) are associated with some maps of the circle
$S_{\infty\text{ }}^{1}$extendable (quasi conformally) to the interior of the
disc \textbf{D} as discussed in our earlier work Ref.[3] and Section 5. An
image of the closed geodesic on $\mathcal{S}$, when lifted to \textbf{H}%
$^{2},$ is just a segment of a circle whose both ends lie on $S_{\infty}^{1}$.
Since \ geodesics are nonintersecting, circle segments on $S_{\infty}^{1}$ are
also nonintersecting. In order to recover results of W-K model, in this work
we are interested only in the $\mathit{periodic}$ maps of the circle as it
will be explained in Section 4 (after Eq.(4.4)). In connection with such maps
the following remark is of importance.

\textbf{Remark 3.3.} (A variant of Sarkovskii theorem, Ref.[79], page 88) Let
$f$: $S^{1}\rightarrow S^{1}$ be a continuous map of the circle with a
periodic orbit of period 3. If the lift $\tilde{f}$: $\mathbf{R}%
\rightarrow\mathbf{R}$ has also a periodic orbit of period 3 then, $f$ has
periodic orbits of every period. The condition on the lift \ map \~{f} cannot
be dropped. In Section 5 we argue that \ even though in the case of W-K model
the continuous maps of the circle are to be replaced by the piecevise linear
maps still 3 remains as minimal period.

\textbf{Remark 3.4.} As noted by Kontsevich [9], moduli space problem makes
sense only for Riemann surfaces obeying the following set of inequalities
\begin{equation}
g\geq0,n>0,2-2g-n<0 \tag{3.4}%
\end{equation}
with $n$ being the number of distinct marked points (effectively distinct
boundary components). Boundary components can be eliminated by the
\textit{Schottky double} construction. This construction can be performed as
follows. If $M$ is a complex manifold with $C_{1},...,C_{n}$ boundary
components, one can consider an exact duplicate of it, say $\hat{M},$ with the
same number of boundary components, say, $\hat{C}_{1},...,\hat{C}_{n}%
.$\ Evidently, for each point $x\in M$ there is a ''symmetric'' point $\hat
{x}\in\hat{M}$ . The Schottky double $2M$ is formed as a disjoint union
$M\cup\hat{M}$ and identifying each point $x\in C_{i}$ with point $\hat{x}%
\in\hat{C}_{i}$ for 1$\leq i\leq n.$ In the simplest case we have initially
either punctured torus, i.e.$g=1,n=1$, or the trice punctured sphere, i.e.
$g=0,n=3$. In both cases the Schottky double is a double torus. A double torus
has 3 geodesics which belong to the geodesic lamination $\mathcal{L}$. The
image of these geodesics lifted to \textbf{H}$^{2}$ produces 3 circular arcs
whose ends lie on $S_{\infty}^{1}$. This is minimal number of arcs required
for the moduli space problem to make sense. According to Remark 3.3. this is
also a minimal period\ for the periodic homeomorphisms of the circle in view
of the Sarkovskii theorem. More on this topic will be discussed in Section 5.

In the mean time we would like to discuss the general case of Riemann surfaces
of genus $g\geq1$ with $n$ boundary components. It is argued in Ref.[80] that
the total number of geodesics on the Schottky double is $6g-6+3n$. This is the
dimension of space of holomorphic quadratic differentials (real on each of the
boundary components). Hence, in accordance with Teichm\"{u}ller theory [20],
it is the dimension of the Teichm\"{u}ller and, accordingly, the moduli space
of such Schottky doubled surface.

\textbf{Remark 3.5. }The dimension of moduli space of Schottky doubled surface
coincides with the dimension of moduli space $\mathcal{M}_{g,n}^{comb}$ of
3-valent ribbon graphs \ used \ in Kontsevich paper, Ref.[9]. In the present
case the same dimension for the moduli space (as obtained by \ Kontsevich) is
obtained without explicit use of the quadratic differentials. Moreover, the
equivalent of the Kontsevich-Penner ribbon graphs are to be obtained below in
Section 4.

It is convenient to map the circle at infinity $S_{\infty}^{1}$ into the real
axis \textbf{R.} Accordingly, the arcs corresponding to closed geodesics on
the Schottky doubled Riemann surface will become semicircles whose ends are
located on the real axis \textbf{R}. This is depicted in Fig.3.%
\begin{figure}
[ptb]
\begin{center}
\includegraphics[
trim=0.000000in -0.018616in 0.000000in 0.018615in,
natheight=0.875200in,
natwidth=8.802100in,
height=0.5172in,
width=4.9606in
]%
{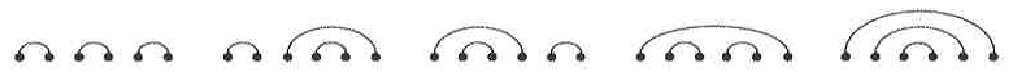}%
\caption{Combinatorics of Catalan numbers presented through arrangements of
non intersecting arcs (representing closed hyperbolic geodesics in the upper
half plane Poincare model of H$^{2})$}%
\end{center}
\end{figure}

The mathematical problem associated with arrangement of the arcs depicted in
Fig.3 can be formulated according to Stanley [42] as follows: It is required
to find a number of ways to connect $2n$ points in the plane lying on a
horizontal line by $n$ nonintersecting arcs, each arc connecting two of the
points and lying above the points. The solution of this problem is just the
Catalan number, $C_{n}=\frac{1}{n+1}\left(
\begin{array}
[c]{c}%
2n\\
n
\end{array}
\right)  $. It was used before in connection with construction of the
partition function for\ our 2+1 gravity model [1,2]. Catalan numbers are very
helpful in solving the mathematical problem of enumeration of meanders.
Meanders had been used as well for description of the partition function of
2+1 gravity [1] and other useful statistical mechanical, dynamical and
biological models [81,82]. Meanders can be easily constructed from the double
set of arcs as illustrated in Fig.4 taken from our earlier work, Ref.[1].

\subsection{Partition function of 2+1 gravity}

W-K treatment of 2 dimensional topological quantum gravity is done in the
grand canonical formalism. This means that such treatment requires two steps.
First, one should construct the (canonical) partition function for the fixed
genus Riemann surface. Second, one should perform summation over all genera
with some chemical potential. Evidently, the second step can be performed only
if the results of the first step are available. For this reason in our
previous works [1,2,4] only the first step was considered. Since the main
Kontsevich identity, Eq.(1.35), is written for the canonical (fixed genus)
case, it is sufficient to consider the fixed genus case in this work also. In
order to reach an agreement with W-K results, it is necessary to reconsider
our earlier obtained results for the canonical partition function of 2+1 gravity.

Let us begin with some reminders. Meander of order $n$ is a closed
nonselfintersecting curve which intersects some straight line in exactly $2n$
\ preassigned points. In Ref.[1] we had discussed the way meander can be
constructed from 2 arc systems, e.g. like those depicted in Fig.3. For
reader's convenience we reproduce a fragment of such construction in Fig.4.%
\begin{figure}
[ptb]
\begin{center}
\includegraphics[
natheight=2.802000in,
natwidth=9.844200in,
height=1.4321in,
width=4.9632in
]%
{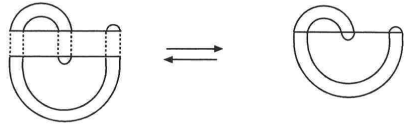}%
\caption{Construction of a typical meander}%
\end{center}
\end{figure}
It is clear from this figure that, in general, such procedure of constructing
meanders will yield a set of disconnected meanders. For a given fixed number
$n$ let the number $M_{n}^{(k)}$ denote the total number of \ disconnected
\ topologically distinct meanders whose total number is $k$. It is clear, that
1$\leq k\leq n $, and that
\begin{equation}
C_{n}\leq M_{n}\leq C_{n}^{2} \tag{3.5}%
\end{equation}
where $M_{n}=M_{n}^{(k=1)}.$ Each meander configuration has some statistical
weight $x$=$\exp\{-\beta J\}$ (with $\beta$ being some fictitious inverse
temperature and J is related to the surface energy) dictated by the physics of
the problem [1,4] so that the total canonical partition function $Z(x)$ is
given by
\begin{equation}
Z_{g}(x)=\sum\limits_{n=0}^{\infty}x^{n}\sum\limits_{k=1}^{n}M_{n}^{(k)}g^{k}
\tag{3.6}%
\end{equation}
with $g$ being determined implicitly through the equation
\begin{equation}
<k>=g\frac{\partial}{\partial g}\ln Z_{g}(x) \tag{3.7}%
\end{equation}
with $<k>$ denoting the average number of meanders in the cluster $k.$This
number is expected to be assigned. If this is not the case, the partition
\ function should be written differently.

\textbf{Remark 3.6}. The partition function $Z_{g}(x)$ is written for the
system of meanders forming a measured foliation on a Riemann surface
$\mathcal{S}$ of fixed genus $g$. Being guided by the Proposition 3.2., we
would like to consider the lift of such foliation to the universal cover of
$\mathcal{S}$ , i.e. to \ the unit disc model of \textbf{H}$^{2}$.

\textbf{Remark 3.7.} In both Ref.[1] and Ref.[4] the main interest in
obtaining the partition function was to study dynamical transition from the
pseudo-Anosov ( hyperbolic) to periodic (Seifert- fibered) regime. In the
present case only the periodic(i.e.Seifert fibered) regime is studied. This is
explained in Section 4. Seifert fibered regime is \textbf{not} discussed in
Refs.[1,4]. \textit{Only in this ( periodic) regime} \textit{W-K \ results can
be recovered from 2+1 gravity}.

To construct the partition function on the unit disc, it is convenient to map
the unit disc into the upper half plane model for \textbf{H}$^{2}$. Then, the
system of arcs representing geodesics on $\mathcal{S}$ is mapped into that
depicted in Fig.3. Next, the arc system depicted in Fig.3 \ is mapped \ into
the associated random walk as depicted in Fig.5.
\begin{figure}
[ptb]
\begin{center}
\includegraphics[
natheight=4.875000in,
natwidth=7.760800in,
height=2.3947in,
width=3.7965in
]%
{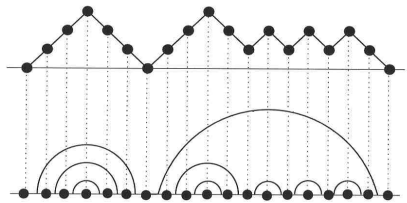}%
\caption{A typical Dyck path [42] from (0,0) to (2n,0) made of steps (1,1) and
(1,-1) never falling below x-axis is in one to one correspondence with a
typical arrangement of arcs.}%
\end{center}
\end{figure}
Because of such mapping, it becomes possible to connect the combinatorics of
vicious walkers discussed in Section 2 with that of arcs and meanders. Indeed,
the walk depicted in Fig.5 is directed and is not allowed to intersect $x$
axis, except at initial and final points. By analogy with Fig.4 one can think
about another Dyck walk (below $x$ axis). Since both walks intersect each
other only at initial and final points this situation looks almost the same as
for two vicious walkers. Following Labelle [83], we can make it identical to
that for the vicious walkers. To achieve this goal, it is sufficient to
translate the upper part by the vector (1,1) while the lower part by the
vector (1,-1) so that one obtains either a problem about statistics of one
vicious walker in the presence of the absorbing wall or about statistics of
two vicious walkers. Both problems are discussed in Fisher's paper [52] and
are actually equivalent. Hence, results of Section 2 can be applied now and
one obtains the Tracy-Widom kernel, Eq.(2.18), in the asymptotic limit of
large genus or large number of boundary components. This is obviously not
sufficient. To go beyond this simple minded result requires to make several
nontrivial mappings (bijections). We shall be brief \ in describing these
bijections since details of the proofs can be found in the published literature.

We begin with the observation that to each Dyck path, e.g. that in Fig.5, one
can associate the Dyck word so that the Dyck path having length 2n is encoded
by a Dyck word of length 2n. The word is composed of letters x and \={x} in
such a way that each North-East (respectively South-East) step correspond to
the letter x (respectively \={x}). The peaks (respectively troughs) correspond
to the factors x\={x} (respectively \={x}x). Instead of North-East
(respectively South-East) steps it is possible to choose strictly North
(respectively East) steps for the Dyck paths to get a configuration like that
depicted in Fig.2. Hence, we obtain the following set of bijections: a) from
the set of arcs to the set of Dyck paths; b) from set of Dyck paths to the set
of Dyck words; c) from the set of Dyck words to the set of lattice paths from
$(0,0)$ to $(n,n)$ with steps $(0,1)$ or $(1,0)$ never rising above the line
$x=y$ ( incidentally one can construct instead the lattice paths which never
go below the diagonal $x=y$ [84] ).To this set of bijections we need to add
two more now. The first one is between the Dyck path, e.g. that depicted in
Fig.5, and the parallelogram polyomino. Polyomino can be made out of squares
(called \textit{cells}). A finite connected union of cells such that the
interior is also connected and there are no cut points is called
\textit{parallelogram polyomino}. It is defined with accuracy up to
translation. As depicted in Fig.6,%
\begin{figure}
[ptb]
\begin{center}
\includegraphics[
natheight=6.156600in,
natwidth=5.843500in,
height=2.3938in,
width=2.2745in
]%
{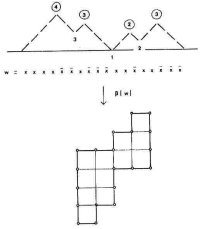}%
\caption{Bijections between the Dyck path, the Dyck word and parallelogram
polyomino}%
\end{center}
\end{figure}
a parallelogram polyomino is bordered by two non-intersecting paths having
only North and East steps. Fig.6 depicts bijections between the Dyck path, the
Dyck word w and parallelogram polyomino $\beta($w). In such bijection the
magnitude and the order of peaks and troughs determine the shape of the
polyomino [85]. Parallelogram polyomino is in one to one correspondence with
skew Young (or Ferres) diagram of shape $\lambda/\mu$ . In the example
displayed in Fig.6 we have the shape (4,4,4,2,2,1)/(3,2). That is from
standard looking Young table of shape (4,4,4,2,2,1) a piece in the upper left
corner is truncated which is also standard table of shape (3,2). Hence, the
skew Young diagrams differ very inessentially from standard looking Young
diagrams. Using this circumstance, we need to exhibit yet another bijection.
It is the most important for our development. To this purpose we need to use
the Theorem 1.11. and Corollary 1.12. in order to state yet another

\bigskip

\textbf{Theorem 3.8. }There is a weight -preserving bijection $\varphi$
between non-intersecting paths (P$_{1},...,$P$_{k})$ and column strict Young
tableaux of shape $\lambda$ with entries from N.

\bigskip

To demonstrate that this is indeed the case, we follow the example discussed
in Ref.[51] .More general case of skew column strict Young diagram is
discussed in Ref.[42](section 7.16). Following Ref.[51] we take $k=4$,
$\lambda=(5,3,2,2)$ and N=6. Although we have now 4 nonintersecting paths
their trajectories are completely specified by labeling of their horizontal
edges (y-coordinates), e.g. see Fig.2. The information about the path
P$_{k+1-i}$ \ \ can be placed (encoded) into the row $i$ of tableau T. In our
case the (vicious) paths are depicted in Fig.7%
\begin{figure}
[ptb]
\begin{center}
\includegraphics[
natheight=5.124900in,
natwidth=8.239900in,
height=2.3947in,
width=3.8346in
]%
{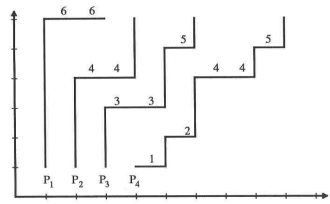}%
\caption{Four vicious walkers encoded by their horizontal edges}%
\end{center}
\end{figure}
while the corresponding encoding of the Young table is depicted in Fig.8.
Thus, using results of Section 1, especially Theorem 1.11 and Corollary 1.12,
we obtain the determinant det M which contains all the configurational
information about collection of vicious walkers.%
\begin{figure}
[ptbptb]
\begin{center}
\includegraphics[
natheight=3.427200in,
natwidth=4.280800in,
height=2.3947in,
width=2.9862in
]%
{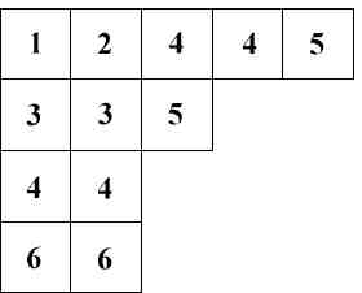}%
\caption{The Young tableau which encodes the information about trajectories of
vicious walkers}%
\end{center}
\end{figure}
Using results of Section 2 this information is being translated into that for
$n$-point correlation functions $R_{n},$ Eq.(2.24), and $T_{n},$ Eq.(2.17).
Using this information and by analogy with Eq.(3.6) the partition function can
be written in the following tentative form:
\begin{equation}
Z[\mathbf{x}]=\sum\limits_{n=1}^{\infty}\int\prod\limits_{i=1}^{n}dy_{i}%
\exp(\mathbf{x}\cdot\mathbf{y})T_{n}[\mathbf{y}]\equiv\sum\limits_{n=1}%
^{\infty}Z_{n}(\mathbf{x}), \tag{3.8}%
\end{equation}
where, the boldface for $x$ and $y$ variables reflects the fact that they are
multidimensional, e.g. \textbf{x}$\cdot$\textbf{y}=$\sum\nolimits_{i=1}%
^{n}x_{i}y_{i},etc.$ We use the word ''tentative'' because we would like at
this point to make a connection with recent works by Okounkov [77,86] and
Okounkov and Pandharipande [87] who use the asymptotics of Hurwitz numbers to
arrive at results similar to our Eq. (3.8). For reasons which will become
clear in the next section and for the sake of agreement with these recent
works, Eq.(3.8) should be rewritten as follows:
\begin{equation}
Z_{n}[\mathbf{x}]=Z[x_{1},...,x_{n}]=\frac{(-1)^{n+1}\left(  2\pi\right)
^{\frac{n}{2}}}{\prod\limits_{i}\sqrt{x_{i}}}\mathcal{E}(\frac{\mathbf{x}%
}{2^{\frac{1}{3}}}) \tag{3.9}%
\end{equation}
with
\begin{equation}
\mathcal{E}(\mathbf{x})=\int\prod\limits_{i}^{n}dy_{i}\exp(\mathbf{x}%
\cdot\mathbf{y})T_{n}(\mathbf{y}). \tag{3.10}%
\end{equation}
\textbf{Remark 3.9.} For developments presented in this paper Eq.s (3.9) and
(3.10) are actually unnecessary. They are provided, nevertheless, because they
are interesting in their own right and, although the initial arguments in
these papers are different from ours, the end results, e.g. Eq.(3.10), is in
accord with our Eq.(3.8).

The interest in these equations stems from the fact that in the large $n$
limit, Eq.(3.10) acquires familiar in physics literature path integral form.
This fact may lead to some new physical applications. Already many physical
applications of vicious walker models can be found in papers by Fisher [52]
and \ by Huse and Fisher [53].

In the present case, using results of Tracy and Widom ( Eq.(4.6) of Ref.[56])
Eq.(2.18) of Section 2 can be rewritten as follows
\begin{equation}
K(X,Y)=\int\limits_{0}^{\infty}dzAi(X+z)Ai(z+Y) \tag{3.11}%
\end{equation}
with Airy function defined in Eq.(2.19). Using Okounkov's Lemma 2.6, Ref.
[77],
\begin{equation}
\int\limits_{-\infty}^{\infty}dzAi(z+a)Ai(z+b)\exp(xz)=\frac{1}{2\sqrt{\pi x}%
}\exp(\frac{x^{3}}{12}-\frac{a+b}{2}x-\frac{\left(  a-b\right)  ^{2}}{4x}),
\tag{3.12}%
\end{equation}
Eq.(3.10) can be rewritten as follows:
\begin{equation}
\mathcal{E}(\mathbf{x})=\frac{1}{2^{n}\pi^{\frac{n}{2}}}\frac{\exp\left(
\frac{1}{12}\sum\nolimits_{i}^{n}x_{i}^{3}\right)  }{\prod\limits_{i}%
\sqrt{x_{i}}}\int\limits_{0}^{\infty}\prod\limits_{i}^{n}dy_{i}\exp
(-\sum\limits_{i=1}^{n}\frac{(y_{i}-y_{i+1})^{2}}{4x_{i}}-\sum\limits_{i=1}%
^{n}\frac{y_{i}+y_{i+1}}{2}x_{i}), \tag{3.13}%
\end{equation}
provided that the cyclic boundary condition , $y_{i+1}=y_{1},$ is imposed.
This makes the above ''path integral'' reminiscent to that for the ring
polymer near hard nonpenetrable wall in the presence of some stretching force [67].

To make connections with results of Kontsevich [9] we need to rewrite the
partition function $Z[\mathbf{x}]$ as genus expansion:
\begin{equation}
Z[\mathbf{x}]=\sum\limits_{g=0}^{\infty}Z_{g}(\mathbf{x}) \tag{3.14}%
\end{equation}
where, in view of Eq.(1.2),
\begin{equation}
Z_{g}(\mathbf{x})=\sum\limits_{\Sigma d_{i}=3g-3+n}<\tau_{d_{1}}\cdot
\cdot\cdot\tau_{d_{n}}>_{g}\prod\limits_{i=1}^{n}x_{i}^{d_{i}}. \tag{3.15}%
\end{equation}
The question arises immediately about connections of Eq.s (3.14) and (3.15)
with Eq.s (3.8)-(3.10). This issue is addressed and treated in Refs.[77,87]
and, therefore, there is no need to repeat the arguments presented in these
references here. Instead, to reach essentially the same goals we would like to
use different arguments in this work.

In Section 1.1. we had noticed that the calculation of tautological averages
is effectively equivalent to the calculation of the Weil-Petersson volumes of
$\mathcal{\bar{M}}_{g,n}.$ In algebraic geometry there is a Wirtinger formula
for such \ volume calculations [12]. In connection with Grassmannians and
Schubert calculus this formula had been discussed in the fundamental paper by
Chern [46]. Wirtinger-like formula is used in Kontsevich paper as well
(e.g.see Section 3 of Ref.[9]).Hence, the partition function of
two-dimensional topological gravity is effectively the generating function for
the Weil-Petersson volumes. In the fixed genus case such volume according to
Kontsevich is given by
\begin{equation}
vol_{p^{\ast}}(\mathcal{\bar{M}}_{g,n})=\frac{1}{d!}\int\limits_{\mathcal{\bar
{M}}_{g,n}}(p_{1}^{2}c_{1}(\lambda_{1})+\cdot\cdot\cdot+p_{n}^{2}c(\lambda
_{n}))^{d} \tag{3.16a}%
\end{equation}
with $c_{1}$($\lambda_{i})$ being the first Chern class of the $i-$th line
bundle, $i=1,...,n$ and $d=3g-3+n$. The above expression \ becomes a true
volume when in the set $p^{\ast}$ of indeterminates $p^{\ast}=$($p_{1}%
,...,p_{n})$ (arbitrary sequence of positive numbers) each $p_{i}$ is being
put equal to one. This is not necessary, however, since the quantity of
interest is the product given by Eq.(1.2) which is obtainable anyway with
$p_{i}^{^{\prime}}$s $\ $being different from one$.$ Hence, the indeterminates
actually play a role of an auxiliary variables analogous to $x_{i}$ in
Eq.(3.15). The connection with Schur polynomials and, hence, with tau function
is clear if one combines Eq.(1.28) and Porteous formula, Eq.(1.27) with
Eq.(3.16a). Finally, arguments \ presented in Section 2, especially,
Eq.(2.36), provide needed justification of Eq.(3.15) since the result,
Eq.(3.16), of Kontsevich can be equivalently rewritten as
\begin{equation}
vol_{p^{\ast}}(\mathcal{\bar{M}}_{g,n})=sgn\times\sum\limits_{\Sigma d_{i}%
=d}<\tau_{d_{1}}...\tau_{d_{n}}>\prod\limits_{i=1}^{n}\frac{p_{i}^{2d_{i}}%
}{d_{i}!}. \tag{3.16b}%
\end{equation}
To make connection with the matrix models additional steps are required. For
instance, Kontsevich is making a Laplace transform of Eq.(3.16b) in order to
obtain
\begin{equation}
\mathcal{L(}vol_{p^{\ast}}(\mathcal{\bar{M}}_{g,n}))(\lambda_{1}%
,...,\lambda_{n})=\sum\limits_{\Sigma d_{i}=d}<\tau_{d_{1}}...\tau_{d_{n}%
}>\prod\limits_{i=1}^{n}\frac{\left(  2d_{i}\right)  !}{d_{i}!}\lambda
_{i}^{-\left(  2d_{i}+1\right)  } \tag{3.17}%
\end{equation}
with $\lambda_{i}$ being the Laplace variable conjugate to $p_{i}$. Taking
into account that $(2d_{i})!/d_{i}!=2^{d_{i}}\cdot\left(  2d_{i}-1\right)  !!$
the overall factor of $2^{d}$ drops out from the main Kontsevich identity,
Eq.(1.35). In view of the results of Refs.[77,86,87] and that going to be
presented in Section 4, in order to reach an accord with the result of
Kontsevich, Eq.(3.17), it is also necessary to perform the Laplace transform
on $Z_{g}(\mathbf{x})$ in Eq.(3.15) provided that this expression is properly
rescaled. The Laplace transform is obtained then as follows:
\begin{equation}
\mathcal{L}\left(  Z_{g}(\mathbf{x})\right)  (\xi_{1},...,\xi_{n}%
)=\int\limits_{0}^{\infty}\prod\limits_{i=1}^{n}dx_{i}\exp(-\mathbf{\xi}%
\cdot\mathbf{x})Z_{g}(\mathbf{x})\frac{1}{\left(  2\pi\right)  ^{\frac{n}{2}%
}\prod\nolimits_{i=1}^{n}x_{i}^{\frac{1}{2}}}. \tag{3.18}%
\end{equation}
Application of the formula
\begin{equation}
\int\limits_{0}^{\infty}dyy^{n-\frac{1}{2}}\exp(-sy)=\Gamma(n+\frac{1}%
{2})s^{-n-\frac{1}{2}}=\sqrt{2\pi}\frac{\left(  2n-1\right)  !!}{2^{n+\frac
{1}{2}}s^{n+\frac{1}{2}}} \tag{3.19}%
\end{equation}
to Eq.(3.18) produces the expected result:
\begin{equation}
\mathcal{L}\left(  Z_{g}(\mathbf{x})\right)  (\xi_{1},...,\xi_{n}%
)=\sum\limits_{\Sigma d_{i}=d}<\tau_{d_{1}}...\tau_{d_{n}}>\prod
\limits_{i=1}^{n}\frac{\left(  2d_{i}-1\right)  !!}{\left(  2\xi_{i}\right)
^{d_{i}+\frac{1}{2}}}, \tag{3.20}%
\end{equation}
where, in order to achieve an agreement with Kontsevich, one needs to make
identification: $\lambda_{i}=\sqrt{2\xi_{i}}$ in Eq.(1.35). Clearly, such
identification is ultimately connected with rescaling made in Eq.(3.18).The
justification of this rescaling is explained in the next section. This is also
needed for completion of the proof of Kontsevich identity, Eq.(1.35) of
Section 1.

\section{Ribbon graphs,Young tableaux and Kontsevich identity}

\subsection{Construction of ribbon graphs}

Ribbon graphs had been invented by Penner [ 62] ( similar construction can be
found also in papers by Harer [88] ) for description of the moduli space
$\mathcal{M}_{g,n}$ of Riemann surfaces. In physics literature similar
construction had been independently developed by Saadi and Zwiebach [63] and
later developed in numerous papers by Zwiebach, e.g. see Ref. [89] cited in
Kontsevich work, Ref.[9]. Both Zwiebach and Kontsevich use the Jenkins-Strebel
quadratic differentials (e.g. see Section 3 of \ our earlier work, Ref.[2],
for condensed summary of their properties) for construction of the ribbon
graphs. In this section we would like to develop alternative method of
construction of ribbon graphs which does not require explicit use of quadratic
differentials. We would like to explain the rationale for constructing the
ribbon graphs in connection with results obtained thus far in this paper.
First, the ribbon graphs appear naturally in matrix models for strings [90]
and QCD [91]. Hence, they are physically interesting and easily constructible
using Feynman-like rules known in quantum field theory [92]. Second,
mathematically these graphs are interesting for several reasons : a) they are
associated with problem of imbedding of, say, trivalent graphs into Riemann
\ surface of fixed genus [93] and, since graphs are related to combinatorial
group theory [94], this problem is also of group-theoretic interest; b) these
graphs are also related to train tracks discussed in our work, Ref.[2].
Because of this connection with train tracks, the number of potential
practical applications is expected to be well beyond particle physics as
explained in the discussion section of Ref.[2]. Third, and most important for
the purposes of this work, they are needed for establishing one-to one
correspondence between the topological gravity and the random matrix models of
string theory. The aspects of this correspondence are discussed in detail in
Ref.[90]. Mathematically, this correspondence is reflected in the Kontsevich
identity, Eq.(1.35). Hence, following Kontsevich [9] it is necessary to prove
that the r.h.s.of Eq.(1.35) is equal to the l.h.s. In sections 2 and 3 we had
demonstrated that the l.h.s of the identity, Eq.(1.35), is associated with the
enumeration of \ allowed configurations for an assembly of vicious walkers.
This problem has been mapped into enumeration of the Young tableau, e.g. see
Figs 7 and 8. These pictures provide a geometrical way of describing
partitions of \ nonnegative integers. Hence, for ribbon graphs we have to find
analogous partitions. Evidently, the Kontsevich identity, Eq.(1.35), is just
the statement about existence of different but equivalent ways of \ describing
partitions as had been stated in the Introduction.

To describe partitions associated with ribbon graphs we need to construct such
graphs first. To facilitate reader's understanding, we employ some results on
graphs and ribbon graphs from the pedagogically written paper by Mulase and
Penkava [64].Ultimately,our way of constructing the ribbon graphs is different
from that discussed in this reference since we are not using quadratic differentials.

\textbf{Definition 4.1.} A graph $\Gamma$=($\mathcal{V}$, $\mathcal{E}$ ; i)
consists of a finite set of vertices $\mathcal{V}$=($\mathit{V}_{1}%
,...,\mathit{V}_{m})$ and finite set of edges $\mathcal{E}$ together with a
map i from $\mathcal{E}$ to the set ($\mathcal{V}\times\mathcal{V}$)/S$_{2}$
of unordered pairs of vertices called incidence relation. The quantity
\begin{equation}
a_{ij}=\left|  i^{-1}(V_{i},V_{j})\right|  \tag{4.1}%
\end{equation}
is the number of edges connecting vertices $V_{i}$ and $V_{j}$ . The degree
(valence) of a vertex $V_{j\text{ }}$ is the number
\begin{equation}
\deg(V_{j})=\sum\limits_{k\neq j}a_{jk}+2a_{jj} \tag{4.2}%
\end{equation}
which is the number of edges incident to the vertex.

\textbf{Definition 4.2.} A graph isomorphism is a pair ($\alpha,\beta)$ of
bijective maps $\alpha:\mathcal{V}\rightarrow\mathcal{V}^{\prime}$and
$\beta:\mathcal{E}\rightarrow\mathcal{E}^{\prime}$ that preserve the incidence relation.

To construct a ribbon graph the above definitions should be modified. The
modification consists in labeling of middle points of each edge thus
effectively creating extra degree 2 vertices associated with each edge. We
denote this extra vertex set as \ $\mathcal{V}_{\mathcal{E}}.$ Now the new set
of vertices is the disjoint union $\mathcal{V}\sqcup\mathcal{V}_{\mathcal{E}}$
while the new set of edges is the disjoint union $\mathcal{E}\sqcup
\mathcal{E}$ since the midpoint of each edge now divides it into two parts.
The incidence relation is described now by the map
\begin{equation}
i_{\mathcal{E}}:\mathcal{E}\sqcup\mathcal{E}\longrightarrow\mathcal{V}%
\times\mathcal{V}_{\mathcal{E}} \tag{4.3}%
\end{equation}
because each edge of $\Gamma_{\mathcal{E}}$ is connecting one vertex of
$\mathcal{V}$ to one vertex of $\mathcal{V}_{\mathcal{E}}$ .Obviously, an edge
of $\Gamma_{\mathcal{E}}$ is called a \textit{half edge} of $\Gamma.$ For
every vertex \textit{V}$\in\mathcal{V}$ of $\Gamma$ , the set i$_{\mathcal{E}%
}^{-1}(\{V\}\times V_{\mathcal{E}})$ consists of half edges incident to
\textit{V }so that
\begin{equation}
\deg(V)=\left|  i_{\mathcal{E}}^{-1}(\{V\}\times V_{\mathcal{E}})\right|
\tag{4.4}%
\end{equation}

\textbf{Definition 4.3}. A ribbon graph is a graph $\Gamma$=($\mathcal{V}$,
$\mathcal{E}$; i) together with a cyclic ordering on the set of half-edges as
depicted in Fig.9.%

\begin{figure}
[ptb]
\begin{center}
\includegraphics[
natheight=4.875000in,
natwidth=11.250400in,
height=2.1646in,
width=4.9606in
]%
{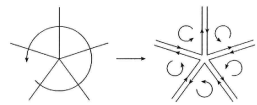}%
\caption{A cyclic ordering at the vertex of the ribbon graph as compared with
no ordering in the case of a vertex for ordinary graph}%
\end{center}
\end{figure}

Such definition of the ribbon graph leads to the following construction of
such graph out of ordered vertices. The strips corresponding to the two half
edges are connected following the orientation of their boundaries to form
ribbons. The final surface is no longer planar in general. It is an oriented
surface whose boundaries are made of boundaries of the ribbons. To illustrate
this construction, let us consider the simplest case first. To this purpose,
we recall that for the case of punctured torus and trice punctured sphere we
have 3 geodesics (on the Schottky doubled surface) which we lift to the
Poincare disc. We associate with these geodesics a trivalent graph as depicted
in Fig.10.
\begin{figure}
[ptb]
\begin{center}
\includegraphics[
trim=0.000000in -0.010350in 0.000000in 0.010351in,
natheight=1.916400in,
natwidth=2.603900in,
height=1.9545in,
width=2.6455in
]%
{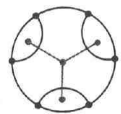}%
\caption{An elementary 3 valent vertex associated with geodesics lifted to the
Poincare disc}%
\end{center}
\end{figure}
\ 

Next, we make another copy of this picture. Then, we thicken the edges
emanating from the vertices and provide orientation in accord with Fig.9.
Next, we glue the strips to each other. This can be done in two ways. One is
depicted in Fig.11%
\begin{figure}
[ptb]
\begin{center}
\includegraphics[
natheight=3.749800in,
natwidth=10.499700in,
height=1.7902in,
width=4.9632in
]%
{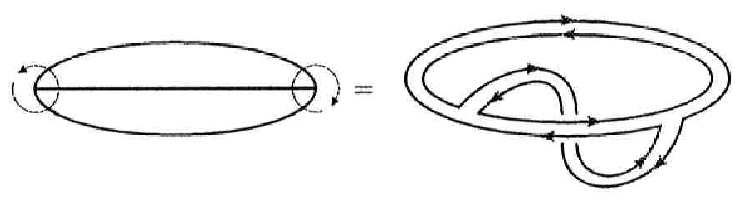}%
\caption{One way to make the simplest ribbion graph (the underlying ''normal''
graph is shown for comparison) which is topologically equivalent to the
punctured torus. Notice, that the same underlying ''normal'' graph also gives
rise to another ribbon graph which is topologically equivalent to the trice
punctured sphere (not to be confused with that depicted in Fig.12).}%
\end{center}
\end{figure}
while the other is depicted in Fig.12.%
\begin{figure}
[ptbptb]
\begin{center}
\includegraphics[
natheight=1.208100in,
natwidth=7.344000in,
height=0.8397in,
width=4.9614in
]%
{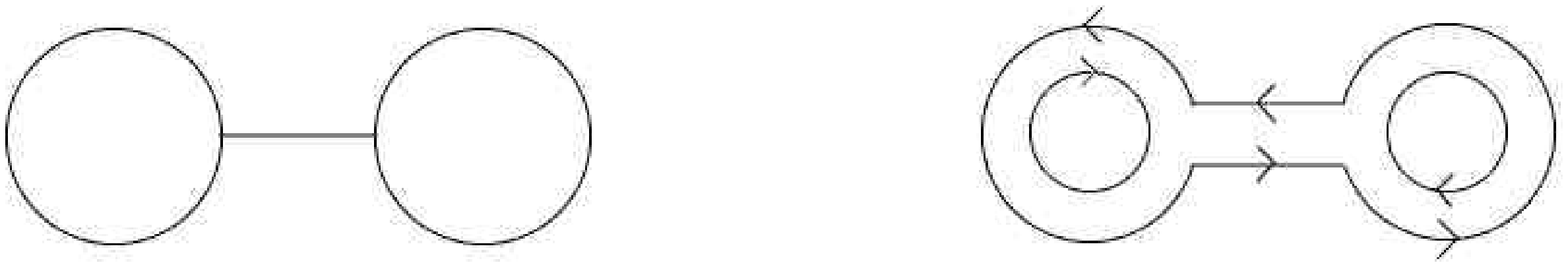}%
\caption{Another way of making the simplest ribbon graph (the underlying
''normal'' graph is also shown for comparison). The ribbon graph is
topologically eqivalent to the trice punctured sphere}%
\end{center}
\end{figure}
We would like to notice that in both cases we have constructed ribbon graphs
$\Gamma_{g,n}=\Gamma$ with the number of thickened edges equal to $6g-6+3n$
number of edges (e.g. 3 for both the trice punctured sphere and the punctured
torus) in accord with Kontsevich [9].To obtain more complicated graphs it is
convenient to proceed by induction. To this purpose, following again Ref.[64],
we need to define the operations of \textit{contraction} and
\textit{expansion}.

\textbf{Definition 4.4.} If the edge E of $\Gamma$ is incident to two distinct
vertices V$_{1}$ and V$_{2}$ another ribbon graph $\Gamma^{\prime} $ called
\textit{contraction} of $\Gamma$ is obtained from $\Gamma$ by removing the
edge E and joining the vertices V$_{1}$ and V$_{2}$ to a single vertex with
the cyclic ordering at the joint vertex determined by the cyclic order of the
edges incident to V$_{1}$starting with the edge following E up to the edge
preceding E, followed by the edges incident to V$_{2\text{ }}$starting with
the edge following E and ending with the edge preceding E.

Evidently, the contraction procedure decreases the number of edges and
vertices by one. Every ribbon graph can be obtained from the trivalent ribbon
graph by applying a sequence of contractions.

\textbf{Definition 4.5}. The expansion of the ribbon graph is operation
\ \textit{inverse} to the contraction.

This means that every trivalent ribbon graph can be constructed by the
expansion procedure starting from a very simple analogue of Fig.10. In view of
the Euler equation
\begin{equation}
\mathcal{V}(\Gamma)-\mathcal{E}(\Gamma)=2-2g-n \tag{4.5}%
\end{equation}
such procedure will produce the desired Kontsevich-type trivalent ribbon
graph. Let us consider this construction in some detail. For example some
representative contraction and expansion of an edge \ is depicted in Fig.13 .
\begin{figure}
[ptb]
\begin{center}
\includegraphics[
natheight=6.562200in,
natwidth=8.593600in,
height=2.3938in,
width=3.1272in
]%
{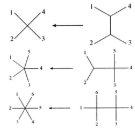}%
\caption{Typical contraction-expansion (Whitehead) moves characteristic for
W-K model}%
\end{center}
\end{figure}

\textbf{Remark 4.6}. The reader familiar with our earlier work, Ref.[1], can
easily recognize the Whitehead moves characteristic for train tracks. This
fact alone is sufficient for making connection between ribbon graphs and
quadratic differentials. Since in the present case no topology changing moves
are involved, this justifies our earlier statement that the results of W-K can
be recovered from the Seifert-fibered (periodic) phase of 2+1 gravity.

For any vertex of degree $d\geq4$ there are $d(d-3)/2$ ways of expanding it by
adding an edge. Consider a portion of $\Gamma$ \ made of vertex of degree
$d\geq4$ and $d$ half edges emanating from this vertex. A \textit{dual} to
this portion is a convex polygon of $d$ sides depicted in Fig.14 along with
the underlying vertex.
\begin{figure}
[ptb]
\begin{center}
\includegraphics[
natheight=3.749800in,
natwidth=15.374600in,
height=1.2306in,
width=4.9614in
]%
{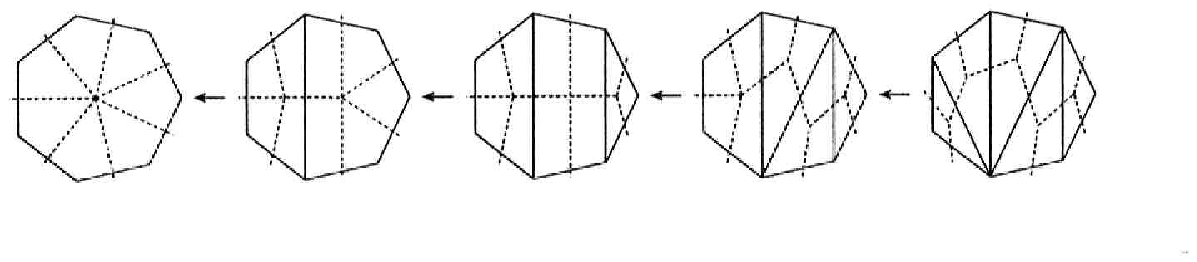}%
\caption{Expansion/contraction of a representative vertex facilitated by the
associated with it dual polygon}%
\end{center}
\end{figure}
The series of contractions/expansions is ultimately related to the number of
ways a convex polygon with $d$ sides can be triangulated by nonintersecting
diagonals. This number is Catalan number again [42]. Although one may probably
use this observation to develop bijections analogous to those discussed in
Section 3, we leave such an opportunity outside the scope of this work in view
of another options to be discussed below.

To finish our construction of the ribbon graphs we have to keep in mind few
additional facts. First, by looking at Fig.10 and realizing that two copies of
the disc model of \textbf{H}$^{2}$ are needed for construction of the ribbon
graph, it is clear that the process of expansion and gluing of strips to each
other should be allowed to be different in each disc. Hence, we have to
associate the probability of 1/2 to every vertex on each disc. This means that
we have to have an overall (assembly) factor of 2$^{-\mathcal{V}(\Gamma)}$
\ for a particular realization of the ribbon graph. Second, the rest of the
automorphisms of the ribbon graph are, evidently, the same as for the
''normal'' graph.

This concludes our description of the ribbon graph construction.

\subsection{Young tableaux and Kontsevich identity}

In section 3 the Laplace transform, Eq.(3.20), of the partition function for
an assembly of vicious walkers has been obtained. Now it is time to explain
how these vicious walkers reemerge with help of the ribbon graphs. To this
purpose let us notice that the boundary components of these graphs can be
looked upon as made of polygons as Kontsevich had noted in section 2.2. of
Ref.[9]. Hence, each ribbon graph is can be classified by certain \ non
negative integer number of triangles (n$_{3})$, quadrangles (n$_{4})$,
pentagons (n$_{5})$, hexagons (n$_{6})$, etc. Each edge $E_{i}$($\Gamma)$ of
the ribbon graph has some length $l_{i}$. Evidently, these lengths can be
grouped into sets identified with polygons so that the numbers above represent
the multiplicities for these sets. Suppose, that the total length $\textsl{L}$
of all polygons is prescribed in advance. Then, given ribbon graph can be
viewed as particular realization of the partition of \textsl{L} into numbers
associated with lengths of these polygons. Alternatively, one can assign the
total number of polygons (that is the total number of faces n in Eq.(4.5)) and
consider partition of this number into n$_{3}$, n$_{5}$, n$_{6\text{ }},$ etc.
In any case, this means that the Young tableaux can be associated with such
partition and, in view of the results of Section 3, again, the vicious walkers
can be linked with such Young/Ferrers tables. The question remains: how to
connect the vicious walkers with ribbon polygons explicitly? To this purpose
recall Remark 2.1. and discussion which follows. According to this remark and
the following discussion both correlation functions $R_{n}$ and $T_{n}$ are
defined in fact for vicious walks made out of loops. Topologically, our
polygons are also loops. Hence, it makes sense to identify the loops entering
into expressions for $R_{n}$ and $T_{n}$ with those coming from polygons
associated with ribbon graphs. This identification requires some care (that is
it requires some proofs) and, hence, cannot be made straightforwardly as we
would like to demonstrate now.

If we associate (replace) the polygonal paths by the paths of vicious walkers
then, except at vertices, only ''binary interactions'' between these walkers
need to be considered. These ''binary interactions'' are of geometrical origin
since they are effectively equivalent to considering just one vicious walker
in the presence of an absorbing wall [52]. According to Fisher [52], this
means that ''no walk can penetrate the wall and any walk attempting to do so
is eliminated''. In addition to the geometrical constraint on such walk, the
present case differs from that discussed by Fisher because one has to
demonstrate that the walk which survived encounter with one wall entirely
forgets about this encounter when it is facing the next wall. Fortunately,
this is the case. Indeed, the non normalized distribution function
$Q_{n}(x;\sigma)$ for the walk starting at $x=a$ and after $n$ steps reaching
the point $x$ is obtained by Fisher, e.g. see Eq.(5.5) of Ref.[52], and is
given by
\begin{equation}
Q_{n}(x;\sigma)\approx\frac{e^{-\sigma n}}{\sqrt{\frac{\pi}{2}}}\frac
{ax\exp(-\frac{x^{2}}{2n})}{n^{\frac{3}{2}}}, \tag{4.6}%
\end{equation}
where, apart from the factor $aexp(-\sigma n),$one can easily recognize the
probability of the first passage through $x$ at ''time'' n for the random walk
which had started at the origin: $x=0$ [75].This probability has yet another
interpretation more suitable for the problem we are discussing. Indeed,
following Feller, Ref.[75],Chr3, the same expression describes probability for
two dimensional (\textbf{directed}) random walkers which had begun their walk
at the origin and after $n$ steps had ended at some point $x>0,y>0$ \ of
$x,y$- plane, provided that these walkers never cross the $x-$ axis.
Evidently, one may associate $x-$axis with ''time'' (in our case n) direction
while $y$ -axis with ''space'' direction in order to get vicious walk
interpretation of such random walk. In any event, it should be clear that in
space direction (that is perpendicular to the wall) such walk is acting as
random and, hence, it ''forgets'' its past. This observation provides
justification for representation of the ''circular'' vicious walk by a
\ sequence of \textbf{independent} random vicious walkers- \textbf{one} for
\textbf{each} edge of the closed polygon.

Next, we have to find an analogue of the partition function, Eq.(3.8).This is
accomplished in several steps. First, we notice that Eq.(3.8) does not contain
information about the number of steps in the walk, only about the number of
walks. Using Eq.s (3.8) and (4.6) we introduce \textbf{for each edge} the
following generating function (known as perimeter measure [86,87] )
\begin{equation}
C(\xi_{1},\xi_{2})=\int\limits_{0}^{\infty}dxP(\xi_{1},\xi_{2};x) \tag{4.7}%
\end{equation}
where
\begin{equation}
P(\xi_{1},\xi_{2};x)=\int\limits_{0}^{\infty}dn_{1}\int\limits_{0}^{\infty
}dn_{2}Q_{n_{1}}(x;\xi_{1})Q_{n_{2}}(x;\xi_{2}) \tag{4.8}%
\end{equation}
and we put $a=1$ in $Q_{n}(x;\xi)$ in Eq.(4.6). This is permissible since the
factor $exp(-\sigma n)$ can be always amended to absorb $a$ (e.g. read
Fisher's paper, Ref.[ ] ). Using Eqs (4.6)-(4.8) the explicit form of the
perimeter measure can be easily obtained using standard tables of the Laplace
transform. The final result is given by
\begin{equation}
C(\xi_{1},\xi_{2})=\frac{2}{\sqrt{2\xi_{1}}+\sqrt{2\xi_{2}}}. \tag{4.9}%
\end{equation}
This result should be applied to all edges of the ribbon graph and the final
result should take into account all automorphisms $\left|  Aut\Gamma\right|  $
of the underlying ''normal'' trivalent graphs. Taking into account the
assembly factor of 2$^{-\mathcal{V}(\Gamma)}$ the final result can be written
as follows
\begin{equation}
\mathcal{L}\left(  Z_{g}(\mathbf{x})\right)  (\xi_{1},...,\xi_{n}%
)=\sum\limits_{G\in\Gamma_{g,n}}\frac{2^{-\mathcal{V}(\Gamma)}2^{\mathcal{E}%
(\Gamma)}}{\left|  Aut(G)\right|  }\prod\limits_{e\in\mathcal{E}(G)}\frac
{1}{\sqrt{2\xi_{i(e)}}+\sqrt{2\xi_{j(e)}}} \tag{4.10}%
\end{equation}
with indices $i$ and $j$ referring to $i$-th and $j-$th polygons sharing the
same edge.

Since the l.h.s. is given by Eq.(3.20), Eq.(4.10) coincides with Kontsevich
identity, Eq.(1.35), provided that identification $\lambda_{i}=\sqrt{2\xi_{i}%
}$ is made. Since for trivalent graphs $\left|  \mathcal{V}\right|  =\frac
{2}{3}\left|  \mathcal{E}\right|  $ , we obtain
\begin{equation}
2^{-\mathcal{V}(\Gamma)}2^{\mathcal{E}(\Gamma)}=2^{2g-2+n} \tag{4.11}%
\end{equation}
in accord with Ref.[87].

\section{Piecewise linear homeomorphisms of the circle and KdV}

In this section we would like to provide another interpretation to the results
obtained in previous sections. This interpretation is desirable since its aim
is to explain to what extent W-K model is universal from the point of view of
dynamical systems theory. This is also desirable for the purpose of
\ connecting results of this paper with WDVV\ equations and Frobenius
manifolds [27,95]. \ To begin, we need to introduce some information about the
Thompson groups.

\subsection{Some facts about the Thompson groups}

The groups $F$, $T$ and $V$ were introduced by Richard Thompson in 1965.
Unfortunately, his results had not been published. This, nevertheless, had not
stopped line of research initiated by Thompson as can be seen from the
pedagogically written review article by Cannon, Floyd and Parry [96]. Below,
we provide some basic facts on Thompson groups using mainly this reference.
Additional sources will be used whenever they are needed to suit our needs.

Let $F$ be the set of piecewise linear homeomorphisms from the closed unit
interval [0,1] to itself that are differentiable except at finitely many
points. Let $f\in F$ and let $0=x_{0}<x_{1}\cdot\cdot\cdot<x_{n}=1$ be the set
of points at which $f$ is not differentiable. This partition determines
intervals [$x_{i-1},x_{i}]$ for $i=1,...,n$ which are called intervals of the
partition. A partition of [0,1] is called \textit{standard dyadic} partition
if and only if the intervals of partition are standard dyadic intervals.

\textbf{Definition 5.1.} A standard dyadic interval in [0,1] is an interval of
the form [$\frac{a}{2^{n}},\frac{a+1}{2^{n}}]$ where $a$ and $n$ are
nonnegative integers with $a\leq2^{n}-1.$

It is useful to associate a finite ordered rooted binary tree associated with
standard dyadic intervals as depicted in Fig.15.%
\begin{figure}
[ptb]
\begin{center}
\includegraphics[
natheight=3.958300in,
natwidth=7.083700in,
height=2.3947in,
width=4.2627in
]%
{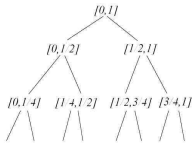}%
\caption{The tree of standard dyadic intervals}%
\end{center}
\end{figure}
For $x_{i-1}\leq x\leq x_{i},i=1,...n,$ the function $f$ can be written as
follows
\begin{equation}
f(x)=a_{i}x+b_{i} \tag{5.1}%
\end{equation}
with $a_{i}$ being a power of 2 and $b_{i}$ being a dyadic rational number. It
can be shown that $f^{-1}\in F$ and $f$ maps the set of dyadic rational
numbers bijectively to itself. Hence, $F$ is closed under the composition of
functions and therefore is a subgroup of the group of all homeomorphisms from
[0,1] to [0,1]. This group is $F$ group of Thompson.

\textbf{Definition} \textbf{5.2}. When points 0 and 1 are identified to make a
circle $S^{1}$, then, the resulting Thompson group is called $T$ group.
Another Thompson group acting on the circle is $V$ group. Its definition is a
bit technical [96] but, at ''physical level'' of rigor, the difference between
$V$ and $T$ groups is hardly noticeable. Therefore, following Ref.[97] we
shall denote both $V$ and $T$ groups as PL$_{2}(S^{1})$ and we shall keep in
mind that both groups are subgroups of the group $Homeo_{+}$ of piecewise
orientation-preserving homeomorphisms of the circle. The following proposition
proven in Ref.[96] is very important.

\bigskip

\textbf{Proposition 5.3}. Let $f\in F$. Then there exists a standard dyadic
partition $0=x_{0}<x_{1}\cdot\cdot\cdot<x_{n}=1$ such that $f$ is linear on
every interval of the partition and $0=f(x_{0})<f(x_{1})\cdot\cdot
\cdot<f(x_{n})=1$ is a standard dyadic partition.

\bigskip

Evidently, using the proof of this proposition, the same statements can be
made for Thompson groups $T$ and $V$ .

\textbf{Remark 5.4.}Using results of Ref.[42], the tree depicted in Fig.15 can
be put in bijective correspondence with the set of non intersecting arcs
depicted in Fig.3. In view of Proposition 5.3., different arc configurations
correspond to different piecewise linear homeomorphisms of the circle caused
by PL$_{2}(S^{1})$. This statement is going to be examined further below in
Section 5.3.

\textbf{Remark 5.5}. In view of Eq.(5.1) it can be shown [98] that, except for
points $0=x_{0}<x_{1}\cdot\cdot\cdot<x_{n}=1,$ the group PL$_{2}(S^{1})$ is
isomorphic to group PSL$_{2}(\mathbf{Z})$ which had been discussed at length
in our previous work, Ref.[4].

\subsection{Some facts about the Ptolemy group}

In addition to isomorphism mentioned in Remark 5.5. there is yet another
important isomorphism between the Thompson and the Ptolemy groups which we
would like to describe briefly in this subsection. Following Lochak and
Schneps [99] define the standard marked tessellation of the Poincare disc as
dyadic tessellation with marked (oriented) edge from 0 to $\infty$ as depicted
in Fig.16. \ %

\begin{figure}
[ptb]
\begin{center}
\includegraphics[
natheight=8.250300in,
natwidth=9.375400in,
height=2.3947in,
width=2.7181in
]%
{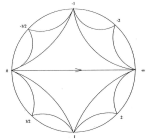}%
\caption{The standard dyadic tesselation of the Poincare disc with oriented
marked edge being displayed}%
\end{center}
\end{figure}
The elementary move $\alpha$ on the oriented edge of the dyadic tessellation
changes its location from one diagonal of the unique quadrilateral containing
it to another one by turning it counterclockwise as depicted in Fig.17.%
\begin{figure}
[ptbptb]
\begin{center}
\includegraphics[
natheight=6.375400in,
natwidth=13.875000in,
height=2.2943in,
width=4.9606in
]%
{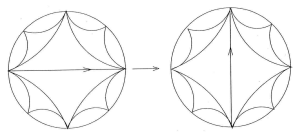}%
\caption{The elementary move on the oriented edge}%
\end{center}
\end{figure}
Such elementary move is of order 4 evidently.\ In addition to this move there
is an arrow-moving move $\beta$ depicted in Fig.18.%
\begin{figure}
[ptbptbptb]
\begin{center}
\includegraphics[
natheight=6.499900in,
natwidth=14.125000in,
height=2.2987in,
width=4.9623in
]%
{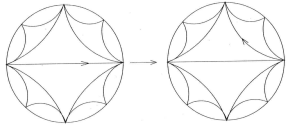}%
\caption{An arrow-moving move}%
\end{center}
\end{figure}
This operation moves an oriented edge to another edge without changing the
tessellation itself. The minimal order of this operation is 3. This fact
\ makes it formally similar to that coming from Sarkovskii theorem, e.g. see
Remark 3.3, which forbids periodic orbits of period lesser than 3.Generators
$\alpha$ and $\beta$ along with relations
\begin{equation}
\alpha^{4}=1;\beta^{3}=1;\left(  \alpha\beta\right)  ^{5}=1 \tag{5.2}%
\end{equation}
and commutator relations
\begin{equation}
\lbrack\beta\alpha\beta,\alpha^{2}\beta\alpha\beta\alpha^{2}]=1;[\beta
\alpha\beta,\alpha^{2}\beta\alpha^{2}\beta\alpha\beta\alpha^{2}\beta^{2}%
\alpha^{2}]=1 \tag{5.3}%
\end{equation}
define the Ptolemy group [99].Imbert had demonstrated [97] that this group is
isomorphic to PL$_{2}(S^{1}).$ The same result is obtained in Ref.[99] with
help of slightly different methods.

\subsection{Circular homeomorphisms, combinatorics of the ribbon graphs and
volumes of hyperbolic polyhedra}

Now we would like to add some important details to Remarks 5.4.and 5.5 made
above. In view of the Remark 5.5. the standard dyadic tessellation of the
Poincare disc depicted in Fig.16 can be shown [97,98] to be in one to one
correspondence with the Farey tessellation of the unit disc \textbf{D}
discussed in detail in Ref.[4]. Such tessellation can be obtained from two
''seeds'', that is from two triangles with vertices located at 0=1/0, 1=1/1 ,
$\infty=1/0$ and 0,-1 and $\infty$ respectively (e.g. see Fig.16 and Section 2
of Ref.[4]) The duals to the tessellations originating from these seeds are
rooted binary trees whose roots are connected to each other as depicted in
Fig.9 of Ref.[4]. The Farey tessellation is also known as \textit{ideal
triangulation} of \textbf{D} [98]. Ideal since the triangle which cover the
hyperbolic plane/disc are ideal in the same sense as Euclidean equilateral
triangles covering the entire plane without gaps form an ideal triangulation
of the Euclidean plane. Since the Farey tessellation is multileveled (e.g.
Fig.16 depicts just two levels while Section 2 of Ref.[4] describes how to
construct tessellations of any level). If the level is finite(or at least
periodic as depicted in Fig.8 of Ref.[4]), then the duals of these
tessellations are two rooted binary trees that are also finite. With help of
Ref.[42] they can be put in bijective correspondence with \textbf{two} systems
of arcs, e.g. those depicted in Figs.3, and 10, etc. needed for construction
of the finite ribbon graph as discussed in Section 4. Surely, the infinite
level case of Farey tessellation is obtained as limiting case of \ finite
level case. The moves depicted in Fig.s 17 and 18 are in fact homeomorphisms
of the circle [98,100] belonging to the subgroup $Homeo_{+}$. They affect the
pattern of triangulation of the unit disc and, hence, affect the binary trees.
As demonstrated in Ref.[96], the group of automorphisms of the reduced binary
trees ( meaning of the word \textit{reduced} is explained in Section 2 of
Ref.[96].) coincides with Thompson group PL$_{2}(S^{1}).$ In the same
reference the analogue of moves $\alpha$ and $\beta$ for trees is obtained.
Some additional physics can be associated now with these mathematical facts.
To this purpose, following Ref.[101] we notice that the automorphisms of two
binary trees associated with two arc systems are independent. Each of these
trees is in bijective correspondence with some triangulation of the convex
polygon by non crossing diagonals [42]. Every automorphism of individual tree
causes some change in the triangulation pattern for given polygon. Since these
changes are independent, following Sleator, Trajan and Thurston [102 ] one may
ask about the minimal number of automorphic steps needed to bring one
triangulation pattern in exact correspondence with another. This problem is of
major importance in computer science. To make it also of importance to physics
(and also to evolutionary genetics [103], dynamics and thermodynamics of
folding of RNA [82,104], etc.) following Ref.[102 ] several more steps are
required. First, by gluing these two polygons to each other we form a
triangulated sphere $S^{2}$. Second, we project this sphere stereographically
into the complex plane so that pattern of triangulations on $S^{2}$ is
transferred to the Euclidean plane. Next, we associate this plane with
boundary of the upper half space model for \textbf{H}$^{3}.$The triangular
pattern on such plane is sufficient for reconstruction of the hyperbolic
polyhedra according to Thurston [16]. A pedagogical account of how it can be
actually done can be found in Ref.[105 ]. Finally, if needed, such polyhedra
can be mapped into the hyperbolic ball model of \textbf{H}$^{3}$ so that the
vertices of these hyperbolic polyhedra are located at the sphere at infinity
$S_{\infty}^{2}.$ Each triangulation pattern on the sphere $S_{\infty}^{2}$
thus associated with some hyperbolic polyhedron of finite volume $V_{H}$ which
is determined by this triangulation pattern. The physics can be now injected
into this picture by introducing some Boltzmann factor exp(-$\beta V_{H})$ so
that different volumes are related to different triangulations. The larger the
volume is for a given ''temperature''$\beta$ , the more stable triangulation
pattern becomes. Incidentally, Ref.[102 ] is devoted to finding hyperbolic
polyhedra with large volumes.

\textbf{Remark 5.6.} Very recently \ statistical mechanics \ of various
physical systems with Boltzmann factor containing volume (including hyperbolic
volume) had been considered by Atiyah and Sutcliffe [106 ]. Actually, these
authors had considered instead of volume its logarithm.

\subsection{Universal Teichm\"{u}ller space, KdV and Frobenius manifolds}

Connections between the Dyck paths, Virasoro characters and exactly solvable
lattice models had been known for some time [107 ].Moreover, since
combinatorics of the Dyck paths is directly associated with that for the
Catalan numbers, such connections are actually not totally unexpected since
there are about 150 interpretations of the Catalan numbers [42]. This, by the
way, means that potentially there are much more applications of the W-K model
than we had mentioned so far. It is not the purpose of \ this subsection to
provide the list of such applications. Instead, we are interested in intrinsic
features of W-K model which are encoded by the Thompson (or the Ptolemy)
group. Following Penner [98,100], the \textit{universal} \textit{Teichm\"{u}%
ller space} $\mathcal{T}_{ess}$ \ \ is defined by
\begin{equation}
\mathcal{T}_{ess}=\frac{Homeo_{+}}{M\ddot{o}b} \tag{5.4}%
\end{equation}
where $M\ddot{o}b$ is just $PSL(2,R)$.If $\mathcal{T}_{ess^{^{\prime}}}$ is a
set of all tessellations of \textbf{D, }then $\mathcal{T}_{ess}=\mathcal{T}%
_{ess^{^{\prime}}}/M\ddot{o}b$ as well, as Penner shows. In Section 7 of
Ref.[3] we had introduced and discussed the universal Teichm\"{u}ller space
T(1) as defined by Bers. More explicitly, such space is defined by
\begin{equation}
\text{T(1)=}\frac{QS}{PSL(2,R)} \tag{5.5}%
\end{equation}
with QS being a set of all quasisymmetric deformations of the circle
$S_{\infty}^{1}$ (or real line \textbf{R}).We shall discuss these deformations
in some detail below. In the meantime we notice that ''physical'' definition
of the Teichm\"{u}ller space T is given by
\begin{equation}
\text{T=}\frac{Diff_{+}}{PSL(2,R)} \tag{5.6}%
\end{equation}
where G= $Diff_{+}$ denote set of all orientation preserving diffeomorphisms
of $S_{\infty}^{1}.$ Since G is proper subgroup of QS [98] , it is clear that
T is embedded into T(1). In Ref.[98] Penner argues that QS is subset of
$Homeo_{+}$ . This leads to the following inclusions:
\begin{equation}
\text{T}\subset\text{T(1)}\subset\mathcal{T}_{ess}. \tag{5.7}%
\end{equation}
This result is very nontrivial since it makes sense out of \ Sarkovskii
theorem (Remark 3.3 and Section 5.2) in the present context. It also allows to
use Nag and Verjovsky [108] arguments (summarized in Section 7 of Ref.[3]) for
use of QS deformations of $S_{\infty}^{1}$ in order to obtain the Virasoro
algebra. It is well documented fact that the method of coadjoint orbits [109]
directly \ connects the Virasoro algebra with KdV equation. The connection
between QS deformations and KdV can be established directly without method of
coadjoint orbits. For this purpose we need to use some classical results of
Ahlfors [65].

Let $w=f(z)$ be a homeomorphism of the complex $z$-plane (or $S^{2}).$Then$,$%
\begin{equation}
dw=f_{z}dz+f_{\bar{z}}d\bar{z}. \tag{5.8}%
\end{equation}
The complex dilatation factor $\mu$ giving rise to the Beltrami equation is
defined now as
\begin{equation}
f_{\bar{z}}=\mu f_{z} \tag{5.9}%
\end{equation}
and the associated modulus of this dilatation factor is defined by
$d_{f}=\left|  \mu\right|  \leq1.$The mapping is considered to be
$K$-\textit{quasiconformal} if there is a nonnegative constant $K$ such that
$D_{f}\leq K$ ,where $D_{f}=\frac{1+d_{f}}{1-d_{f}}.$ Accordingly, the mapping
is conformal if $D_{f}=1.$Suppose now that $f$ maps the upper Poincare half
plane to itself (which is, of course, equivalent to the mapping of \textbf{D}
to itself). The mapping is quasisymmetric, i.e.$f$ $\in QS$ if for all points
$x$, $x-t$ and $x+t$ on real line \textbf{R }the following\textbf{\ }$M$
condition
\begin{equation}
M^{-1}\leq\frac{f(x+t)-f(x)}{f(x)-f(x-t)}\leq M \tag{5.10}%
\end{equation}
is satisfied with $M$ being some non negative constant. In Chr.6 of Ref.[65 ]
Ahlfors proves that the function which is \textit{schlicht} and has a
quasiconformal extension to the upper half plane must obey the following
Fuchsian-type equation known already to Poincare [110]
\begin{equation}
y^{\prime\prime}+\frac{1}{2}\phi y=0. \tag{5.11}%
\end{equation}
The mapping function $f=y_{1}/y_{2}$ where $y_{1}$ and $y_{2}$ are being two
independent solutions of Eq.(5.11) normalized by
\begin{equation}
y_{1}^{\prime}y_{2}-y_{2}^{\prime}y_{1}=1. \tag{5.12}%
\end{equation}
The function $\phi$ is determined by equation $[f]$=$\phi$ with $[f]$ being
the Schwarzian derivative of $f$. Recall [111], that the function is
considered to be schlicht (or simple) at some point of complex plane if its
first derivative is nonzero at this point.

Lazutkin and Pankratova [112 ] studied Eq.(5.11) from the point of view of the
circle maps. They used a variant of Eq.(5.1) given by
\begin{equation}
F(\xi+2\pi)=F(\xi)+2\pi,\xi\in\mathbf{R,} \tag{5.13}%
\end{equation}
to study the transformational properties of Eq.(5.11). In particular, change
of variables
\begin{equation}
x=F(\xi)\text{ and }y\text{(}F\text{(}\xi\text{))=}Y\text{(}\xi)\sqrt{\text{
}F^{\prime}(\xi)} \tag{5.14}%
\end{equation}
leaves Eq.(5.11) in the same form $Y^{\prime\prime}+\frac{1}{2}\Phi Y=0$ with
potential $\Phi$ given by
\begin{equation}
\Phi(\xi)=\phi(F(\xi))\left[  F^{\prime}(\xi)\right]  ^{2}+[F(\xi)] \tag{5.15}%
\end{equation}
with $[F(\xi)],$ again, being the Schwarzian derivative of $F.$ Eq.(5.15)
\ actually determines transformational properties of the Schwarzian
derivative. In \ Ahlfors notations [65] this equation can be equivalently
rewritten as equation for transformation of the Schwarzian derivative:
\begin{equation}
\lbrack f\circ F]=([f]\circ F])\left(  F^{\prime}\right)  ^{2}+[F]. \tag{5.16}%
\end{equation}
Evidently, this equation holds irrespective to the explicit form of
Eq.(5.13).Using this observation it is of interest to consider transformation
of the type $F$($\xi)=\xi+\delta\varphi(\xi)$ with $\delta$ being small
parameter and $\varphi(\xi)$ some function which makes Eq.(5.13) to hold. Use
of this type of function in Eq.(5.16) produces
\begin{equation}
\Phi(\xi+\delta\varphi(\xi))=\phi(\xi)+\delta\left(  \hat{T}\varphi\right)
(\xi)+O(\delta^{2}) \tag{5.17}%
\end{equation}
with
\begin{equation}
\left(  \hat{T}\varphi\right)  (\xi)=\phi(\xi)\varphi^{\prime}(\xi)+\frac
{1}{2}\varphi^{\prime\prime\prime}(\xi)+\phi^{\prime}(\xi)\varphi(\xi).
\tag{5.18}%
\end{equation}
The Korteweg- de Vries equation can be obtained now in a simple minded way. To
this purpose, since $\varphi$ is arbitrary, we choose $\varphi=\phi.$ Next, we
make both $\varphi$ and $\phi$ to depend upon the parameter $\delta$ , that is
we write $\phi=\phi(\xi,\delta).$Next, we assume that the parameter $\delta$
plays role of ''time'' $t$ and, finally, we write
\begin{equation}
\lim_{t\rightarrow0}\frac{\Phi-\phi}{t}=\frac{\partial\phi}{\partial t}.
\tag{5.19}%
\end{equation}
This then produces our final result:
\begin{equation}
\frac{\partial\phi}{\partial t}=3\phi\phi^{\prime}+\frac{1}{2}\phi
^{\prime\prime\prime}. \tag{5.20}%
\end{equation}
Eq.(5.20) coincides with Eq.(2.2) of Segal [113] and, hence, can be called KdV
equation. We provided details of derivation in order to emphasize the
universality of this equation in problems which involve circular maps or maps
of \textbf{D}. Since KdV is effectively dual to the Virasoro algebra surely it
also can be obtained via Nag-Verjovsky approach to construction of the
Virasoro algebra and the Kirillov-Kostant two- form by using the universal
Teichm\"{u}ller space T(1) [108 ] . Summary of Nag-Verjovsky results can be
found in our earlier work, Ref.[3]. These results were extended by Penner [98]
whose Theorem 5.5. essentially assures the existence of the Kirillov-Kostant
two form which is invariant under transformation taken from the Ptolemy group.
Independent and deep studies of the same problem can be found also in earlier
work by Ghys and Sergiescu [114 ]. Higher order KdV can be easily obtained by
sequential use of $\hat{T}$ operator as can be seen from Ref.[115 ].

Eq.(5.11) contains actually much more information than we had discussed thus
far as was shown by Dubrovin, e.g. see Example C.1.of Ref.[95]. From this
example (and preceeding discussion) it follows that variety of \ equations of
Witten-Dijkgraaf-Verlinde-Verlinde (WDVV) type are obtainable as special cases
of Eq.(5.11). Moreover, the self-dual \ Yang-Mills and Einstein equations also
follow from Eq.(5.11) [95,116]. Geometrically, solutions of WDVV\ equations
represent \textit{Frobenius maifolds}. Thy are named after Frobenius who was
the first to discover them in 1882 \ [117]. Incidentally, the self-dual
Einstein equations had been studied already in 1881 by Halphen [118] and
rediscovered by Atiyah and Hitchin [116]. In recent paper by van de Leur and
Martini [119] KP representation theory and the related Sato infinite
Grassmanian are used to construct solutions of the WDVV equations and, hence,
the Frobenius manifolds. In addition, Dubrovin and Zhang [120] had recently
demonstrated that , at least for Frobenius manifolds of genus zero and one the
Virasoro constraints hold true. That is WDVV\ equations can be reduced to that
of KdV type. This result is in complete accord with arguments provided earlier
in this section supporting the claim about universality of KdV type equations.
This universality has its origin in the properties of the universal
Teichm\"{u}ller and moduli space. This universality has been studied
group-theoretically by Schneps and Lochak [99] who studied action of the
Grotendieck-Teichm\"{u}ller group $\hat{G}\hat{T}$ on Ptolemy-Teichm\"{u}ller
groupoid. Recent lecture notes by Bakalov and Kirillov [121 ] contain some
additional helpful information. The authors also discuss connections between
the Catalan numbers, modular functor, \ modular tensor category,
Teichm\"{u}ller tower, Knizhnick-Zamolodchikov equations, etc. and contain
many references on latest important related works. The latest paper by Schneps
and Nakamura [122] should, perhaps, be added to the list references. It is our
hope, that mentioning of all these beautiful mathematical results in this
paper may encourage some applications of these results in areas other than
mathematics and theoretical particle physics.

\textbf{Note added in proof}.While this paper was under refereing, several
important recent publications came to our attention. In particular, Ref.[123]
provides a very efficient introduction into Schubert polynomials, Shubert
varieties and related topics. Refs.[124,125] written and edited by Gian Carlo
Rota provide an indispensable supplement to book by Stanley [42]. These
referenses allow to keep things in historical perspective and, hopefully,
should serve as an inspiration for many additional potential applications. In
recent paper by S\"{o}zen and Bonahon [126] it is shown that the Weil
Petersson symplectic form $\omega_{w-p}$,e.g. see Eq.(1.4), coincides with
Thurston intersection form $\tau$ for geodesic laminations ( more details on
this subject can be found from yet unpublished book by Francis Bonahon
''Closed Curves on Surfaces'' available on line ).This allows to establish
connections between foliations/laminations and non commutative geometry as
actually had been demonstrated by Connes [127] some time ago.In addition to
monograph [27] by Manin, recently Manin and Zograf \ had obtained explicit
asymptotics for Weil-Petersson volumes of the moduli spaces of punctured
Riemann surfaces in the limit of fixed genus and number of punctures groving
to infinity [128] while Grushevsky have obtained the asymtotics for the fixed
number of punctures with genus groving to infinity [129]. In Ref.[130] Guha
made some progress in showing new connections between the diffeomorphisms of
the circle and the exactly integrable systems of the Korteweg-deVries type by
connecting them with differential Galois theory. Finally, direct links between
the results of our Ref. [4] and those discussed in this paper could be also
developed with help of random polynomials [131,132].

\bigskip

\textbf{Acknowledgment}. The autor would like to thank Bojko Bakalov (MIT),
Andrei Okounkov (U. of California, Berkeley) and Igor Rivin (Temple U.) for
stimulating correspondence. In addition, the author would like to thank
Motohico Mulase(U.California, Davis) and Michael Penkava (U.of Wisconsin-Eau
Claire) for granting permission to reproduce Figures \# 9, 11, 13 and 14 in
this work.

\bigskip

\pagebreak 

\bigskip

\ \ \ \ \ \ \ \ \ \ \ \ \ \ \ \ \ \ \ \ \ \ \ \ \ \ \ \ \ \ \ \ \textbf{\ \ \ \ \ \ \ \ \ \ \ \ \ References
}

\bigskip

[1] A.Kholodenko, Use of meanders and train tracks for description of

\ \ \ \ \ defects and textures in liquid crystals and 2+1 gravity,

\ \ \ \ \ J.Geom.Phys.33 (2000) 23-58.

[2] A.Kholodenko, Use of quadratic differentials for description of defects

\ \ \ \ \ \ and textures \ in liquid crystals and 2+1 gravity,

\ \ \ \ \ \ J.Geom.Phys.33 (2000) 59-102.

[3] A.Kholodenko, Boundary conformal field theories, limit sets of Kleinian

\ \ \ \ \ \ groups and holography, J.Geom.Phys.35 (2000) 193-238.

[4] \ A.Kholodenko, Statistical mechanics of 2+1 gravity from Riemann zeta

\ \ \ \ \ \ function and Alexander polynomial:exact results,

\ \ \ \ \ \ J.Geom.Phys.38 (2001) 81-139.

[5] E.Witten, Two-dimensional gravity and intersection theory on moduli

\ \ \ \ \ space, Surveys in Diff.Geom.1 (1991) 243-310.

[6] J.Harris, I.Morrison, Moduli of Curves, Sringer-Verlag,

\ \ \ \ \ New York, 1998.

[7] E.Looijenga,Cellular decompositions of compactified moduli spces of

\ \ \ \ \ pointed curves, in R.Dijkgraaf,C.Faber,G.Vander Geer $\left(
\text{Eds.}\right)  $,

\ \ \ \ \ The Moduli Space of Curves, Birkh\"{a}user, Boston, 1995, pp.369-400.

[8] R.Hain, E.Looijenga, Mapping class groups and moduli spaces of curves,

\ \ \ \ \ Proc.Symp.Pure Math. 62 (1997) 97-142.

[9] M.Kontsevich, Interrsection theory on the moduli space of curves and

\ \ \ \ \ the matrix Airy function,Comm.Math.Phys.147 (1992) 1-23.

[10] Y.Namikawa, A conformal field theory on Riemann surfaces realized

\ \ \ \ \ \ as quantized moduli theory of Riemann surfaces, Proc.Symp.Pure Math.

\ \ \ \ \ \ 49(1) (1989) 413-443.

[11] M.Dunajski, L.Mason, P.Tod, Einstein-Weil geometry, the dKP

\ \ \ \ \ \ \ equation and twistor theory, J.Geom.Phys.37 (2001) 63-93.

[12] P.Griffiths,\ J.Harris, Principles of Algebraic Geometry,

\ \ \ \ \ \ \ John Willey \& Sons, New York, 1994.

[13] V.Danilov, V.Shokurov,\ Algebraic Curves, Algebraic Manifolds

\ \ \ \ \ \ \ \ and Schemes, Springer-Verlag, New York, 1998.

[14] \ C.McMullen, Renormalization and 3-Manifolds which Fiber over the

\ \ \ \ \ \ \ \ Circle, Princeton U.Press, Princeton, 1996.

[15] P.Deligne, D. Mumford, The irreducibility of the space of curves of

\ \ \ \ \ \ \ given genus, Inst. Hautes etudes Sci.Publ.Math. 45 (1969) 75-109.

[16] W.Turston, Geometry and Topology of 3-Manifolds, Princeton

\ \ \ \ \ \ \ University Lecture Notes,1979 http://www.msri.org/gt3m/

[17] W.Thurston, On the geometry and dynamics of diffeomorphisms of

\ \ \ \ \ \ \ surfaces, Bull.AMS (N.S.) 19 (1988) 417-431.

[18] K.Matsutaki, M.Taniguchi, Hyperbolic Manifolds and Kleinian Groups,

\ \ \ \ \ \ \ Clarendon Press, Oxford, 1998.

[19] S.Wolpert, On the homology of the moduli space of stable curves,

\ \ \ \ \ \ Ann.Math.118 (1983) 491-523.

[20] Y.Imayoshi, M.Taniguchi, An Introduction to Teichm\"{u}ller Spaces,

\ \ \ \ \ \ Springer-Verlag, New York, 1992.

[21] E.D'Hoker,D.Phong, The geometry of string perturbation theory,

\ \ \ \ \ \ \ Rev.Mod.Phys.60 (1988) 873-916.

[22] S.Wolpert, On the K\"{a}hler form of the moduli space of once

\ \ \ \ \ \ \ punctured tori, Comm.Math.Helvetici 58 (1983) 246-256.

[23] R.Penner, Weil-Petersson volumes, J.Diff.Geom.35 (1992) 559-608.

[24] P.Zograf, The Weil-Petersson volume of the moduli space of punctured

\ \ \ \ \ \ spheres, Contemp.Math.150 (1993) 367-372.

[25] M.Matone, Nonperturbative model of Liouville gravity, J.Geom.Phys.

\ \ \ \ \ \ 21 (1997) 381-398.

[26] R.Kaufman,Y.Manin, D.Zagier, Higher Weil-Petersson volumes

\ \ \ \ \ \ \ of moduli \ spaces of stable n-pointed curves,

\ \ \ \ \ \ \ Comm.Math.Phys.181 (1996) 763-787.

[27] Y.Manin, Frobenius Manifolds,Quantum Cohomology, and Moduli Spaces,

\ \ \ \ \ \ \ AMS Publications, Providenve, RI, 1999.

[28] R.Wells, Differential Analysis on Complex Manifolds, Prentice Hall Inc.,

\ \ \ \ \ \ \ New York, 1973.

[29] S.Wolpert, On obtaining a positive line bundle from the

\ \ \ \ \ \ \ Weil-Petersson class, Am.J.Math.107 (1985)1485-1507.

[30] S.Wolpert, The hyperbolic metric and the geometry of universal curve,

\ \ \ \ \ \ \ J.Diff.Geom. 31 (1990) 417-472.

[31] H.Sato, Algebraic Topology: An Intuitive approach, AMS Publications,

\ \ \ \ \ \ \ Providence, RI, 1999.

[32] S.Novikov, Topology I, Springer-Verlag, Berlin, 1996.

[33] W.Baily, On the imbedding of V-manifolds in projective space,

\ \ \ \ \ \ \ Am.J.Math.79 (1957) 403-430.

[34] T.Miwa, M.Jimbo, E.Date, Solitons,Cambridge U.Press,

\ \ \ \ \ \ \ Cambridge, 2000.

[35] S.Kleiman, D.Laksov, Schubert calculus, Am.Math.Monthly 79 (1972)

\ \ \ \ \ \ \ 1061-1082.

[36] W.Hodge, D.Pedoe, Methods of Algebraic Geometry, Vol.2,

\ \ \ \ \ \ \ \ Cambridge U.Press, Cambridge, 1952.

[37] D.Gepner, Fusion rings and geometry, Comm.Math.Phys.141

\ \ \ \ \ \ \ (1991) 381-411.

[38] E.Witten, The Verlinde algebra and the cohomology of the Grassmanian,

\ \ \ \ \ \ \ arXiv:hep-th/9312104.

[39] W.Fulton, J.Harris, Representation Theory,

\ \ \ \ \ \ \ Springer-Verlag, Berlin, 1991.

[40] W.Fulton, Young Tableaux, Cambridge U.Press, Cambridge, 1997.

[41] W.Fulton, Intersection Theory, Springer-Verlag, Berlin, 1984.

[42] R.Stanley, Enumerative Combinatorics, Vol.2 , Cambridge U.Press,

\ \ \ \ \ \ \ Cambridge, 1999.

[43] I.Porteous, Simple singularities of maps, LNM192 (1971) 286-312.

[44] G.Horrocks, Proc.London Math.Soc.(3) 7 (1957) 265-280.

[45] J.Carrel, Chern classes of the Grassmanians and Schubert calculus,

\ \ \ \ \ \ \ \ Topology 17 (1978) 177-182.

[46] S.Chern, Characteristic classes of hermitian manifolds,

\ \ \ \ \ \ \ Ann.Math. 47 (1946) 85-121.

[47] C.Ehresmann, Sur la topologie de certains espaces homogenes,

\ \ \ \ \ \ \ \ Ann.Math.35 (1934) 396-443.

[48] T.Suwa, Indices of Vector Fields and Residues of Singular

\ \ \ \ \ \ \ \ Holomorphic Foliations, Hermann, Paris, 1998.

[49] T.Regge,General relativity without coordinates,

\ \ \ \ \ \ \ Nuovo Cim.19 (1961) 558-571.

[50] S.Chern, Complex Manifolds Without Potential Theory,

\ \ \ \ \ \ \ \ Springer-Verlag, Berlin, 1979.

[51] D.Stanton, D.White, Constructive Combinatorics, Springer-Verlag,

\ \ \ \ \ \ \ Berlin, 1986.

[52] M.Fisher, Walks, walls, wetting and melting, J.Stat.Phys.34

\ \ \ \ \ \ (1984) 667-729.

[53] D.Huse, M.Fisher, Commensurate melting, domain walls, and

\ \ \ \ \ \ \ dislocations, Phys.Rev.B 29 (1984) 239-269.

\bigskip\lbrack54] P.Forrester, Exact solution of the lock step model of vicious

\ \ \ \ \ \ \ walkers, J.Phys.A 23 (1990) 1289-1273.

[55] M.Mehta, Random Matrices, Academic Press, New York, 1991.

[56] C.Tracy, H.Widom, Level-spacing distributions and the Airy Kernel,

\ \ \ \ \ \ \ Comm.Math.Phys.159 (1994) 151-174.

[57] P.Forrester, The spectrum edge of random matrix ensembles,

\ \ \ \ \ \ \ Nucl.Phys.B402 (1993) 709-728.

[58] R.Kulkarni,F.Raimond,3-dimensional Lorentz space-forms and

\ \ \ \ \ \ Seifert fiber spaces, J.Diff.Geom.21 (1985) 231-268.

[59] S.Francois, Variete's anti-de Sitter de dimension 3 possedant

\ \ \ \ \ \ \ un champ de Killing non trivial,C.R. Acad.Sci.,

\ \ \ \ \ \ \ Paris,Ser.I Math.324 (1997) 525-530.

[60] W.Dunbar, Geometric orbifolds, Rev.Math.

\ \ \ \ \ \ \ Univ.Madrid 1 (1988) 67-99.

[61] J.Nielsen, \ Untershungen zur Topologie der geslossenen

\ \ \ \ \ \ \ zweiseitigen Flachen I, Acta Math. 50 (1927) 189-358.

[62] R.Penner, The decorated Teichm\"{u}ller space of punctured surfaces,

\ \ \ \ \ \ \ Comm.Math.Phys.113 (1987) 299-339.

[63] M.Saadi, B.Zwiebach, Closed string theory from polyhedra,

\ \ \ \ \ \ \ \ Ann.Phys.192 91989) 213-227.

[64] M.Mulase, M. Penkava, Ribbon graphs, quadratic differentials

\ \ \ \ \ \ \ on Riemann surfaces, and algebraic curves defined over
\textbf{\={Q},}

\ \ \ \ \ \ \ Asian J.Math.2 (1998) 875-919.

[65] L.Ahlfors, Lectures on Quasiconformal Mappings,

\ \ \ \ \ \ \ D.van Nostrand Co.,Inc., New York, 1967.

[66] A.Kholodenko, Fermi-Bose transmutation: from semiflexible polymers

\ \ \ \ \ \ \ \ to superstrings, Ann.Phys.202 (1990) 186-225.

[67] A.Kholodenko,Th.Vilgis, Some geometrical and topological problems

\ \ \ \ \ \ \ \ in polymer physics, Phys. Reports 298 (1998) 251-370.

[68] K.Ito, H.McKean, Diffusion Processes and Their Sample Paths,

\ \ \ \ \ \ \ \ Springer-Verlag, Berlin, 1965.

[69] J.Harlee, Time amd time functions in reparametrized nonrelativistic

\ \ \ \ \ \ \ \ quantum mechanics, Class.Quantum Grav. 13 (1996) 361-375.

[70] M.Gaudin, La Function D'Onde De Bethe, Masson, Paris, 1983.

[71] E.Gutkin, Bethe ansatz and the generalized Yang-Baxter equations,

\ \ \ \ \ \ Ann.Phys.176 (1987) 22-48.

[72] I.Krichever, O.Lipan, P.Weigmann, A.Zabrodin, Quantum

\ \ \ \ \ \ integrable\ models and discrete classical Hirota equations,

\ \ \ \ \ \ Comm.Math.Phys. 188(1997) 267-304.

[73] S.Majid, Foundations of Quantum Group Theory, Cambridge U.Press,

\ \ \ \ \ \ \ Cambridge, 1995.

[74] S.Fomin, A.Kirillov, The Yang-Baxter equation, symmetric functions,

\ \ \ \ \ \ and Schubert polynomials, Discrete Math.153 (1996) 123-143.

[75] W.Feller, An Itroduction to Probabiliy Theory and Its Applications,

\ \ \ \ \ \ \ \ Vol.1, John Wiley \&Sons, Inc., New York, 1968.

[76] F.Dyson, Statistical theory of energy levels of complex systems,

\ \ \ \ \ \ \ J.Math.Phys. 3 (1962) 140-156.

[77] A.Okounkov,Generating functions for intersection numbers on moduli

\ \ \ \ \ \ \ spaces of curves, arXiv: mathAG/0101201.

[78] P.Di Francesco, 2-d quantum and topological gravities,

\ \ \ \ \ \ matrix models, and integrable differential systems, in

\ \ \ \ \ \ The Painleve Property, Springer-Verlag, Berlin, 1999,

\ \ \ \ \ \ pp.229-285.

[79] W. de Melo, S.van Strien, One-Dimensional Dynamics,

\ \ \ \ \ \ \ Springer-Verlag, Berlin, 1993.

[80] F.Tomi, A.Tromba, The index theorem for minimal surfaces of

\ \ \ \ \ \ \ higher genus, AMS Memoirs 117 (1995) 1-78.

[81] P.Di Francesco,O.Golonelli, E.Guttier, Meander, folding and arc

\ \ \ \ \ \ \ statistics, Math.Comput.Modelling 26 (1997) 97-147.

[82] R.Simion, Noncrossing partitions, Discr.Math. 217 (2000) 367-409.

[83] J.Labelle, On pairs of non-crossing generalized Dyck paths,

\ \ \ \ \ \ \ \ J.of Stat.Planning and Inference 34 (1993) 209-217.

[84] J.Fedou, Enumeration of skew Ferres diagrams and basic

\ \ \ \ \ \ Bessel functions, Journ.of Stat.Planning and Inference

\ \ \ \ \ \ 34 (1993) 107-123.

[85] M.Delest, G.Viennot, Algebraic languages and polynomioes

\ \ \ \ \ \ enumeration, Theor.Comp.Sci. 34 (1984) 169-206.

[86] A.Okounkov, Random matrices and random permutations,

\ \ \ \ \ \ Int.Math.Res.Notices 20 (2000) 1043-1095.

[87] A.Okounkov, R.Panharipande, Gromov-Witten theory, Hurwitz

\ \ \ \ \ \ numbers, and matrix models, arXiv: math.AG/0101147.

[88] J.Harer, The cohomology of moduli space of curves,

\ \ \ \ \ \ \ LNM 1337 (1988) 138-221.

[89] B.Zwiebach, How covariant closed string theory solves a minimal area

\ \ \ \ \ \ \ \ problem, Comm.Math.Phys.136 (1991) 83-118.

[90] J.Amjorn, B.Durhuus, T.Josson, Quantum Geometry,

\ \ \ \ \ \ \ Cambridge U.Press,Cambridge, 1997.

[91] W.Taylor, Counting strings and phase transitions in 2d QCD,

\ \ \ \ \ \ \ \ arXiv:hep-th/9404175.

[92] D.Bessis, C.Itzykson, J-B Zuber, Quantum field theory techniques in

\ \ \ \ \ \ \ \ graphical enumeration, Adv.Appl.Math.1 (1980) 109-157.

[93] A.White, Graphs, Groups and Surfaces, North Holland,

\ \ \ \ \ \ \ Amsterdam, 1973.

[94] D.Collins, R.Grigorchuk, P.Kurchanov, H. Zieschang ,

\ \ \ \ \ \ \ Combinatorial Group Theory and Applications to Geometry,

\ \ \ \ \ \ \ \ Springer-Verlag, Berlin, 1998.

[95] B.Dubrovin, Geometry of 2d topological field theories,

\ \ \ \ \ \ \ \ LNM 1620 (1996) 120-348.

[96] J.Cannon,W.Floyd and W.Parry, Introductory notes on Richard

\ \ \ \ \ \ \ Thompson's groups, Enseign.Math.42 (1996)215-256.

[97] \ M.Imbert, Sur l'isomorphisme du groupe de Richard Thompson

\ \ \ \ \ \ \ avec le groupe de Ptolemee', in :Geometric Galois Actions,

\ \ \ \ \ \ \ Cambridge U.Press, Cambridge,1997, pp.313-324.

[98] R.Penner, Universal constructions in Teichm\"{u}ller theory,

\ \ \ \ \ \ \ Adv.Math.98 (1993) 143-215.

[99] \ P.Lochack, L.Schnieps, The universal Ptolemy -Teichm\"{u}ller groupoid,

\ \ \ \ \ \ \ in :Geometric Galois Actions, Cambridge U.Press, Cambridge,

\ \ \ \ \ \ \ 1997, pp. 325-347.

[100] R.Penner, The universal Ptolemy group and its completions,

\ \ \ \ \ \ \ in Geometric Galois Actions, Cambridge U.Press, Cambridge,

\ \ \ \ \ \ \ 1997, pp.293-312.

[101] R.Alperin,W Dicks, J.Porti, The boundary of the Gizeking tree

\ \ \ \ \ \ \ \ in hyperbolic three space, Topology.Appl. 93 (1999) 219-259.

[102] D.Sleator, R.Trajan,W.Thurston, Rotation distance, triangulations

\ \ \ \ \ \ \ \ \ \ and hyperbolic geometry, J.of AMS 1 (1988) 647-681.

[103] M.Yang, Introduction to Statistical Methods in Modern Genetics,

\ \ \ \ \ \ \ \ \ \ Gordon and Breach Publ., 2000.

[104] R.Penner, M.Waterman, Spaces of RNA secondary structures,

\ \ \ \ \ \ \ \ \ \ Adv.Math.101 (1993) 31-49.

[105] P.Callahan, A.Reid, Hyperbolic structures on knot complements,

\ \ \ \ \ \ \ \ \ \ Chaos, solitons and Fractals 9 (1998) 705-738.

[106] M.Atiyah, P.Sutcliffe, The geometry of point particles,

\ \ \ \ \ \ \ \ \ arXiv: hep-th/0105179.

[107] A.Carey, M.Murray, Geometrical Analysis and Lie Theory

\ \ \ \ \ \ \ \ \ in Mathematics and Physics, \ Cambridge U.Press,

\ \ \ \ \ \ \ \ \ Cambridge, 1998.

[108] S.Nag, A.Verjovsky, Diff(S$^{1})$ and the Teichm\"{u}ller spaces,

\ \ \ \ \ \ \ \ \ Comm.Math.Phys.130 (1990) 123-138.

[109] V.Arnold, B.Khesin, Topological Methods in Hydrodynamics,

\ \ \ \ \ \ \ \ \ \ Springer-Verlag, Berlin, 1998.

[110] H.Poincare, Papers on Fuchsian Functions,

\ \ \ \ \ \ \ \ \ \ Springer-Verlag, Berlin, 1985.

[111] H.Cohn, Conformal Mapping on Riemann Surfaces,

\ \ \ \ \ \ \ \ \ \ Dover Publications, Inc., New York, 1967.

[112] V.Lazutkin, T.Pankratova, Normal forms and versal deformations for

\ \ \ \ \ \ \ \ \ \ the Hill equation, Funct.Analysis and Appl. 9 (1975) 41-48.

[113] G.Segal, The geometry of the KdV equation,

\ \ \ \ \ \ \ \ \ \ Int.J.Mod.Phys.A 6 (1991) 2859-2869.

[114] E.Ghys,V.Sergiescu, Sur un groupe remarquable de

\ \ \ \ \ \ \ \ \ diffeomophismes du cercle, comm.Math.Helv. 62 (1987) 185-239.

[115] A.Das, Integrable Models, World Scientific, Singapore, 1989.

[116] M.Atiyah, N.Hitchin, The Geometry and Dynamics of

\ \ \ \ \ \ \ \ \ \ Magnetic Monopoles, Princeton U.Press, Princeton, 1988.

[117] F.Frobenius,Stickelberger, Uber die Differentiation der

\ \ \ \ \ \ \ \ \ elliptishen Functionen nach den Perioden und Invarianten,

\ \ \ \ \ \ \ \ \ J.Reine Angew. Math.92 (1882).

[118] G.Halphen, Sur un systemes d'equations differentielles,

\ \ \ \ \ \ \ \ \ C.R.Acad. Sc. Paris 92 (1881) 1001-1007.

[119] J. van de Leur, R.Martini, The constructionof Frobenius

\ \ \ \ \ \ \ \ \ manifolds from KP tau-functions, Comm.Math.Phys.

\ \ \ \ \ \ \ \ \ 205 (1999) 587-616.

[120] B.Dubrovin, Y.Zhang, Frobenius manifolds and Virasoro constraints,

\ \ \ \ \ \ \ \ \ Sel.Math., New Ser.5 (1999) 423-466.

[121] B.Bakalov, A.Kirillov, Jr., Lectures on Tensor Categories and

\ \ \ \ \ \ \ \ \ Modular Functors, AMS Publ., Providence, R I , 2001.

[122] H.Nakamura, L.Schneps, On subroup of the Grothendieck-Teichm\"{u}ller

\ \ \ \ \ \ \ \ \ group acting on tower of profinite Teichm\"{u}ller modular groups,

\ \ \ \ \ \ \ \ \ Inv.Math. 141 (2000) 503-560.

[123] L.Manivel, Symmetric Functions, Schubert Polynomials and

\ \ \ \ \ \ \ \ Degeneracy Loci, SMF/AMS Texts and Monographs,

\ \ \ \ \ \ \ \ Vol.6, Providence, R.I., 2001.

[124] Gian-Carlo Rota on Combinatorics, Editor J.P.S.Kung, Birkh\"{a}user,

\ \ \ \ \ \ \ \ \ Boston, 1995

[125] Classic Papers in Combinatorics, Edited by I.Gessel and G-C Rota,

\ \ \ \ \ \ \ \ \ Birkh\"{a}user, Boston, 1987.

[126] Y.S\"{o}zen, F.Bonahon,The Weil-Petersson and Thurston symplectic

\ \ \ \ \ \ \ \ \ forms, Duke Math.Journal 108 (2001) 581-597.

[127] A.Connes, A survey of foliations and operator algebras, Proc.Symp.

\ \ \ \ \ \ \ \ \ Pure Math.Vol.38, AMS, Providence, R.I.,1982.

[128] Y.Manin, P.Zograf, Invertible cohomological field theories and

\ \ \ \ \ \ \ \ \ Weil-Petersson volumes, arXiv: math. AG/9902051

[129] \ S.Grushevsky, An explicit upper bound for Weil-Petersson

\ \ \ \ \ \ \ \ volumes of the moduli spaces of punctured Riemann surfaces,

\ \ \ \ \ \ \ \ \ Math.Ann.321( 2001) 1-13.

[130] P.Guha, Diff($S^{1})$ and Adler-Gelfand-Dikii spaces and integrable

\ \ \ \ \ \ \ \ systems. lett.in Math.Phys.55 (2001) 17-31.

[131] A.Edelman, E.Kostlan, How many zeros of a random polynomial

\ \ \ \ \ \ \ \ are real?, Bull.AMS 32 (1995) 1-37.

[132] E.Bogomolny, O.Bohigas,P.Leboefuf, Quantum chaotic dynamics

\ \ \ \ \ \ \ \ \ and random polynomials, J.Stat.Phys. 85 (1996) 639-679.
\end{document}